\documentclass[twocolumn,tighten,usenames,dvipsnames,twocolappendix]{aastex631}
\usepackage[normalem]{ulem}
\usepackage{verbatim}
\usepackage{amssymb}
\usepackage{amsmath}
\usepackage{xspace}
\usepackage{graphicx}
\usepackage{lipsum}
\usepackage{multirow}
\usepackage[T1]{fontenc}
\usepackage[utf8]{inputenc} 

\usepackage{amsmath,mathtools,mathrsfs}
\usepackage{hyperref}
\usepackage{afterpage}
\usepackage{booktabs}
\usepackage{enumitem}

\DeclareUnicodeCharacter{2212}{-}

\usepackage{leftidx}
 \usepackage{txfonts}
 \usepackage{physics}
 \usepackage{amsthm}

\makeatletter
\usepackage{xcolor}

 \shorttitle{Analytic thin disks and rings}
 \shortauthors{Faraji \& Tang}

\begin{document}

%
\title{Analytic thin disks and rings in a class of non-asymptotically flat static spacetimes}


\author[0000-0003-1727-8992]{Shokoufe Faraji}
\affiliation{Department of Physics and Astronomy, University of Waterloo, 200 University Avenue West, Waterloo, ON, N2L 3G1, Canada}
\affiliation{Perimeter Institute for Theoretical Physics, 31 Caroline Street North, Waterloo, ON, N2L 2Y5, Canada}
\affiliation{Center of Applied Space Technology and Microgravity (ZARM), University of Bremen, 28359 Germany}

\author[0000-0000-0000-0000]{Ruijing Tang} 
\affiliation{Observatoire de Paris, Universite PSL, Paris, France}
\affiliation{Perimeter Institute for Theoretical Physics, 31 Caroline Street North, Waterloo, ON, N2L 2Y5, Canada}

\begin{abstract}
External matter distributions can substantially reshape the strong field environment of compact objects, yet this effect is usually neglected in idealized isolated models. In this work, we investigate geometrically thin, optically thick relativistic accretion onto a static axisymmetric space-time that describes a slightly deformed compact object immersed in an external quadrupolar field as an exact solution of vacuum Einstein field equations. Our aim is to determine whether such locally geometries can produce distinctive accretion signatures and, more broadly, to identify the physically meaningful radial domain over which the local solution remains self-consistent. We show that the external quadrupolar distortion leaves a clear imprint on both orbital dynamics and accretion structure. We further find that the outer edge of the radiating region is closely tied to the transition between radiation pressure and gas pressure dominance, which may link the geometry to the thermodynamic properties of the flow. Therefore, the local nature of the distorted spacetime is not merely a formal geometric feature, but has observable consequences for the morphology and emission properties of accretion flows.

\end{abstract}



\section{Introduction}

Relativistic accretion remains one of the sharpest astrophysical probes of the most manifestations of strong gravity. In the thin and radiatively efficient regime, the background spacetime controls the orbital energy and angular momentum of the flow, the location of marginal stability, and the radial distribution of dissipation and emitted flux. For this reason, thin disk theory has long history of the connecting exact relativistic geometries and potentially observable signatures \citep{ShakuraSunyaev1973,NovikovThorne1973,PageThorne1974}. There are vast studies in this research area and these seminal works have been extended in different ways, e.g., \cite{Abramowicz1978RelativisticAccretingDisks,AbramowiczEtAl1988SlimDisks,NarayanYi1994ADAFSelfSimilar,EsinMcClintockNarayan1997States,LiuMeyerMeyerHofmeister1999EvaporationBH,2000ApJ...534..398M,QuataertGruzinov2000CDAF,NarayanIgumenshchevAbramowicz2000CDAF,NarayanIgumenshchevAbramowicz2003MAD,2007ApJ...665.1038C,2014MNRAS.439..503S,2017MNRAS.464.1102B,2019ApJ...880...67J,BegelmanSilk2017MagneticallyElevatedAGN,ChoNarayan2022DiskEvaporation}, among others.

However, most analytical studies still rely on the idealization of an isolated compact object, even though realistic sources are expected to possess both intrinsic multipolar structure and non-negligible environmental influence. In realistic astrophysical settings, compact objects are rarely truly isolated see e.g., \cite{2022PhRvL.129x1103C,2022PhRvD.105f1501C,2023PhRvD.107h4027D,2025PhRvL.135u1403F,2025arXiv250918313S,2025PhRvD.112l4062K,2026PhRvD.113d4040D}.

Their exterior geometry may be shaped by intrinsic multipolar deformations as well as by surrounding matter distributions. Among higher multipole moments, the quadrupole is the leading correction to spherical symmetry and therefore the first that one expects to leave a measurable imprint on orbital motion and accretion structure. Even relatively small quadrupolar distortions can substantially modify the geometry and the associated dynamics in the strong field region (e.g.,  \cite{PhysRevD.24.320,GerochHartle1982,1985ZhETF..88.1921P,2003CQGra..20.5121K,2005GReGr..37.1371H,Quevedo:2010vx,doi:10.1142/S0218271811019852,2012PhRvD..86d4013L,PhysRevD.85.104031,doi:10.1142/S0217751X16410062,PhysRevD.93.024024,2017JCAP...05..039G,2017GrCo...23..149N,PhysRevD.100.044001,2019EPJC...79..730C,2021A&A...654A.100F,2021A&A...654A.100F,2022EPJC...82.1149F,DestounisEtAl2023,2025EPJC...85..148F}). This makes quadrupolarly deformed exact solutions particularly valuable as controlled laboratories for studying how departures from spherical symmetry affect astrophysical processes.

Among the different solutions, the q-metric occupies a distinguished place in the field. It is an exact static, axisymmetric vacuum solution endowed with a quadrupole moment and can be regarded as a simple analytic generalization of Schwarzschild to deformed sources \citep{osti_4201189,doi:10.1063/1.1705005,PhysRevD.2.2119,PhysRevD.39.2904}. Because of its analytic tractability, it has been widely used to study the gravitational field of slightly deformed compact objects and the dynamical consequences of quadrupolar structure.

More recently, the q-metric has been extended to include an additional external static axisymmetric gravitational field, yielding a distorted geometry that combines intrinsic deformation of the source with the influence of an ambient matter distribution \citep{2022Univ....8..195F}. Unlike the seed q-metric, the externally distorted geometry is not globally asymptotically flat and should therefore be interpreted as a local spacetime description. In the present work, we focus on this generalized solution to investigate whether its local character should propagate into the orbital and radiative properties of any accretion flow built on top of it. In that case, the boundary of validity of the spacetime may itself become an astrophysically relevant feature. In fact, even when truncated at the quadrupolar level, the external field can already produce nontrivial modifications in the geometry and in the associated orbital structure that can alter the properties of circular motion, lensing, and other strong gravity observables in nontrivial ways \citep{2021PhRvD.104h3006F,2021Univ....7..447F,2022MNRAS.513.3399F,2023MNRAS.525.1126F,2023mgm..conf..307F,2025PhRvD.111d5003A,2025A&A...699A.266F}. The central idea is that the local character of the distorted spacetime is reflected not only in geodesic observables, but also in the stability, transport, and morphology of accretion disk configurations. By combining these perspectives, we aim to identify a physically motivated radial boundary for the region where the distorted q-metric can be used to clarify which accretion signatures are genuinely associated with the local geometry. This problem is also connected to the broader question of horizonless and naked singularity spacetimes. Independently of whether such configurations are realized in nature, they provide useful theoretical laboratories for disentangling which observational signatures are truly tied to the existence of an event horizon and which instead arise from strong gravity, multipolar structure, or environmental effects e.g, \citep{KovacsHarko2010,JoshiMalafarinaNarayan2014,ShaikhEtAl2019,GyulchevEtAl2020,2026arXiv260215912F}. 

We use two complementary probes of the same underlying question. The first is the dynamics of equatorial circular timelike geodesics, through which we analyze the behavior of the conserved orbital quantities and the structure of marginally stable motion. The second is the relativistic thin disk model, which tests the same background through local shear, radiative transport, pressure transitions, and the morphology of the emitting configuration.

The paper is organized as follows. \autoref{sec:spacetime} presents the background geometry and its main properties. In \autoref{sec:behaviour} we examine the behavior of the physical quantities associated with this spacetime. \autoref{se:stability} discusses the effective potential and the marginally stable circular orbits. In \autoref{sec:diskdomain} we formulate the thin disk model, introduce the relevant equations, and present the corresponding results. Finally, \autoref{sec:summary} contains a summary of the main results and concluding remarks.
\section{Space-time}\label{sec:spacetime}

The q-metric is a static, axisymmetric, and asymptotically flat vacuum solution of the Einstein equations that generalizes the Schwarzschild spacetime through the inclusion of a quadrupole parameter. It represents the exterior gravitational field of an isolated static axisymmetric mass distribution. Starting from the q-metric as a seed solution \citep{PhysRevD.39.2904,2013arXiv1310.5339Q}, one can construct a distorted extension - which is also an exact vaccum solution of Einstein field equation - describing a deformed source in the presence of an external mass distribution \citep{2022Univ....8..195F}. While the seed q-metric is asymptotically flat, the distorted geometry obtained in this way is, in general, only locally valid. In prolate spheroidal coordinates\footnote{Prolate spheroidal coordinates are obtained by rotating the two-dimensional elliptic coordinates about the focal axis of the ellipse.} the corresponding line element takes the form
\begin{align}\label{EImetric}
ds^2 ={}& -\left(\frac{x-1}{x+1}\right)^{1+\alpha} e^{2\hat{\psi}}\, dt^2 \\
&+ M^2 (x^2-1)\left(\frac{x+1}{x-1}\right)^{1+\alpha} e^{-2\hat{\psi}}\nonumber\\
&\quad\Bigg[
\left(\frac{x^2-1}{x^2-y^2}\right)^{\alpha(2+\alpha)} e^{2\hat{\gamma}}
\left(
\frac{dx^2}{x^2-1}+\frac{dy^2}{1-y^2}
\right)
+(1-y^2)\, d\phi^2
\Bigg].\nonumber
\end{align}
where $t \in (-\infty, +\infty)$, $x \in (1, +\infty)$, $y \in [-1,1]$, and $\phi \in [0, 2\pi)$ and $\alpha\in(-1,\infty)$ is the deformation parameter in the q-metric. Here $M$ is the Schwarzschild mass parameter appearing in the metric and equals the physical mass when $\alpha=0$.
For general $\alpha$, $M_{\rm ADM}=(1+\alpha)M$. Since the seed q-metric contribution has already been written explicitly in \autoref{EImetric}, the functions $\hat{\psi}$ and $\hat{\gamma}$ encode only the external distortion field \citep{2022Univ....8..195F}. They are given by
\begin{align}\label{finalpsi}
\hat{\psi}
= \sum_{n=1}^{\infty} \beta_n R^n P_n\!\left(\frac{xy}{R}\right),
\end{align}
and
\begin{align}\label{finalgamma}
\hat{\gamma}
&= \sum_{n=1}^{\infty} \beta_n (1+\alpha)
\sum_{l=0}^{n-1}
\left[(-1)^{\,n-l+1}(x+y)-x+y\right] R^l P_l
\nonumber\\
&\quad
+ \sum_{n,k=1}^{\infty}
\frac{nk\,\beta_n\beta_k}{n+k}
R^{n+k}\left(P_nP_k-P_{n-1}P_{k-1}\right),
\end{align}
where $\beta_n \in \mathbb{R}$ are constants characterizing the external multipolar distortion, $P_n$ denotes the Legendre polynomial of degree $n$, and
\begin{align}
P_n \equiv P_n\!\left(\frac{xy}{R}\right), \qquad
R=\sqrt{x^2+y^2-1}.
\end{align}
By construction, setting $\hat{\psi}=0$ and $\hat{\gamma}=0$ recovers the q-metric, whereas the additional limit $\alpha=0$ yields the Schwarzschild solution. The transformation to Schwarzschild-like coordinates is
\begin{align}\label{transf1}
x=\frac{r}{M}-1,\qquad y=\cos\theta,
\end{align}
so that the domain $x>1$ corresponds to the exterior region $r>2M$.
This metric has a central curvature singularity at $x=-1$ (or $r=0$), as well as an additional singularity appears at $x=1$ (or $r=2M$), and the norm of the time-like Killing vector at this latter vanishes. However, outside this hypersurface, there exists no additional horizon. Nevertheless, considering a relatively small quadrupole moment, a physically interior solution can cover this hypersurface, since it is closely place to the central singularity \citep{Quevedo:2010vx}. Besides, out of this region, there is no more singularity, and the metric is asymptotically flat. 
The spacetime possesses a central curvature singularity at $x=-1$ (equivalently, $r=0$). A second distinguished hypersurface occurs at $x=1$ (or $r=2M$), where the norm of the timelike Killing vector $\xi^\mu\xi_\mu=g_{tt}$ vanishes. This condition, however, is only necessary and not sufficient for the existence of a regular Killing horizon. In the Schwarzschild limit, $x=1$ corresponds to the event horizon, but for a nonvanishing deformation parameter the regularity of this surface is lost and curvature invariants diverge there. For sufficiently small departures from spherical symmetry, the exterior geometry may be matched to a suitable regular interior solution whose boundary lies at a radius larger than $2M$, thereby covering the singular hypersurface and removing it from the physically relevant vacuum domain \citep{Quevedo:2010vx}. In that case, the spacetime can provide an effective description of the exterior field of a slightly deformed compact object. Furthermore, no further singularities arise in the region $x>1$, and the spacetime is asymptotically flat. However, once the external distortion field is included, the resulting geometry is a locally valid rather than globally asymptotically flat \citep{2022Univ....8..195F}.

To isolate the leading effect of the external environment, we truncate the distortion field at the quadrupolar level. This is the natural first nontrivial shape dependent contribution beyond the monopolar sector already encoded in the metric, while any residual dipolar term can be removed by a suitable choice of origin. Accordingly, we set $\beta_n=0$ for $n\neq 2$ and denote
\begin{align}
\beta \equiv \beta_2.
\end{align}
The distortion potentials then reduce to
\begin{align}\label{eq:metricfunctions_xy}
\hat{\psi} &= \frac{\beta}{2}\left(3x^2y^2-x^2-y^2+1\right), \\
\hat{\gamma} &= -2\beta(1+\alpha)x(1-y^2)\nonumber\\
&\quad + \frac{\beta^2}{4}(x^2-1)(1-y^2)\left(-9x^2y^2+x^2+y^2-1\right).
\end{align}
At this level, the distorted q-metric is determined by two parameters aside from the mass scale; the intrinsic deformation parameter $\alpha$, and the external quadrupolar distortion $\beta$. The parameter $\alpha$ measures the deviation of the source from spherical symmetry, and $\beta$ characterizes the strength of the ambient quadrupolar field. In the regime of interest, both $\alpha$ and $\beta$ are taken to be sufficiently small so that the source is only mildly deformed and the external field acts as a weak distortion (more details in \cite{2022Univ....8..195F}).

\begin{figure*}
    \centering 
     \includegraphics[width=0.45\textwidth]{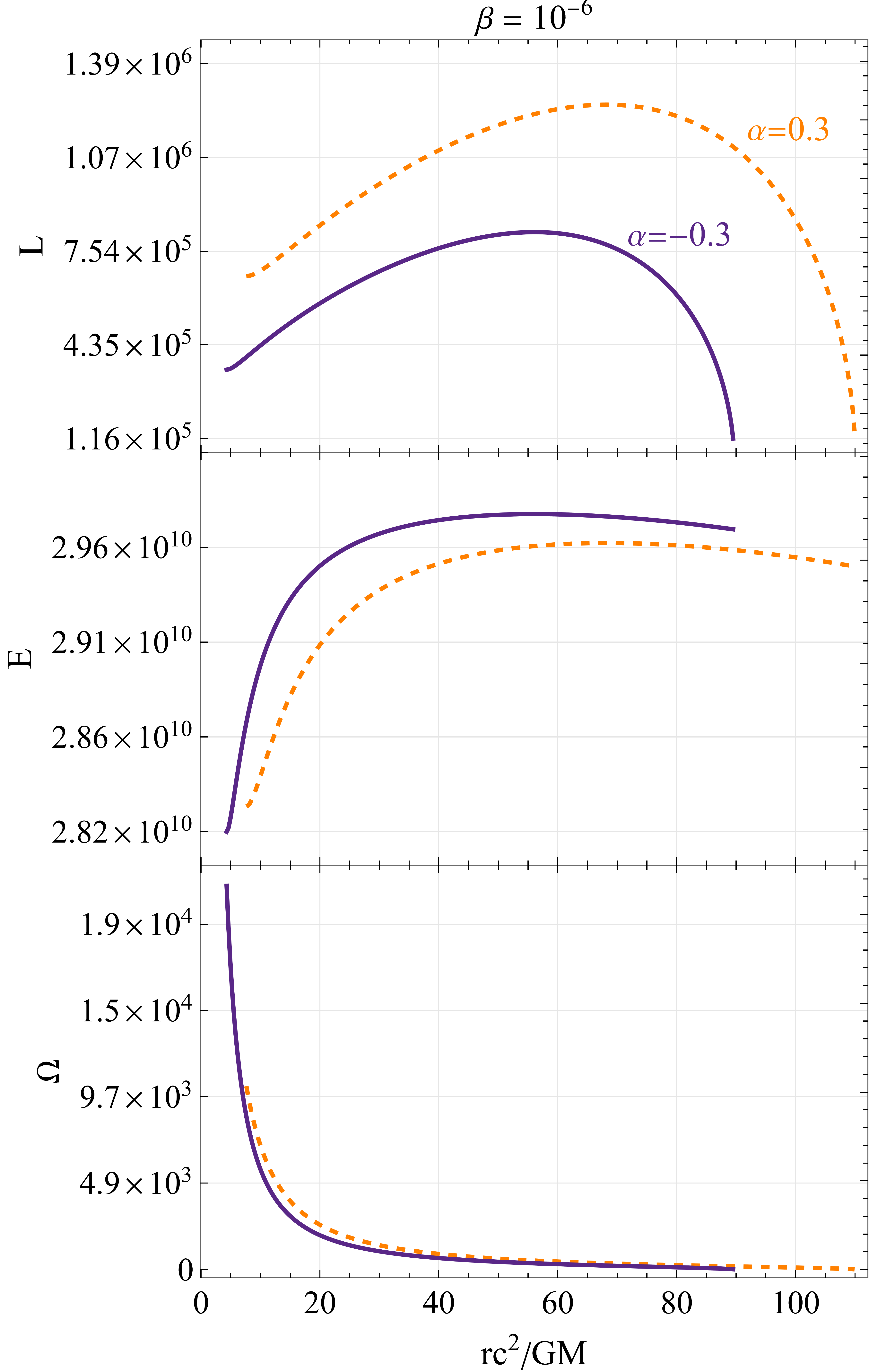}
     \includegraphics[width=0.45\textwidth]{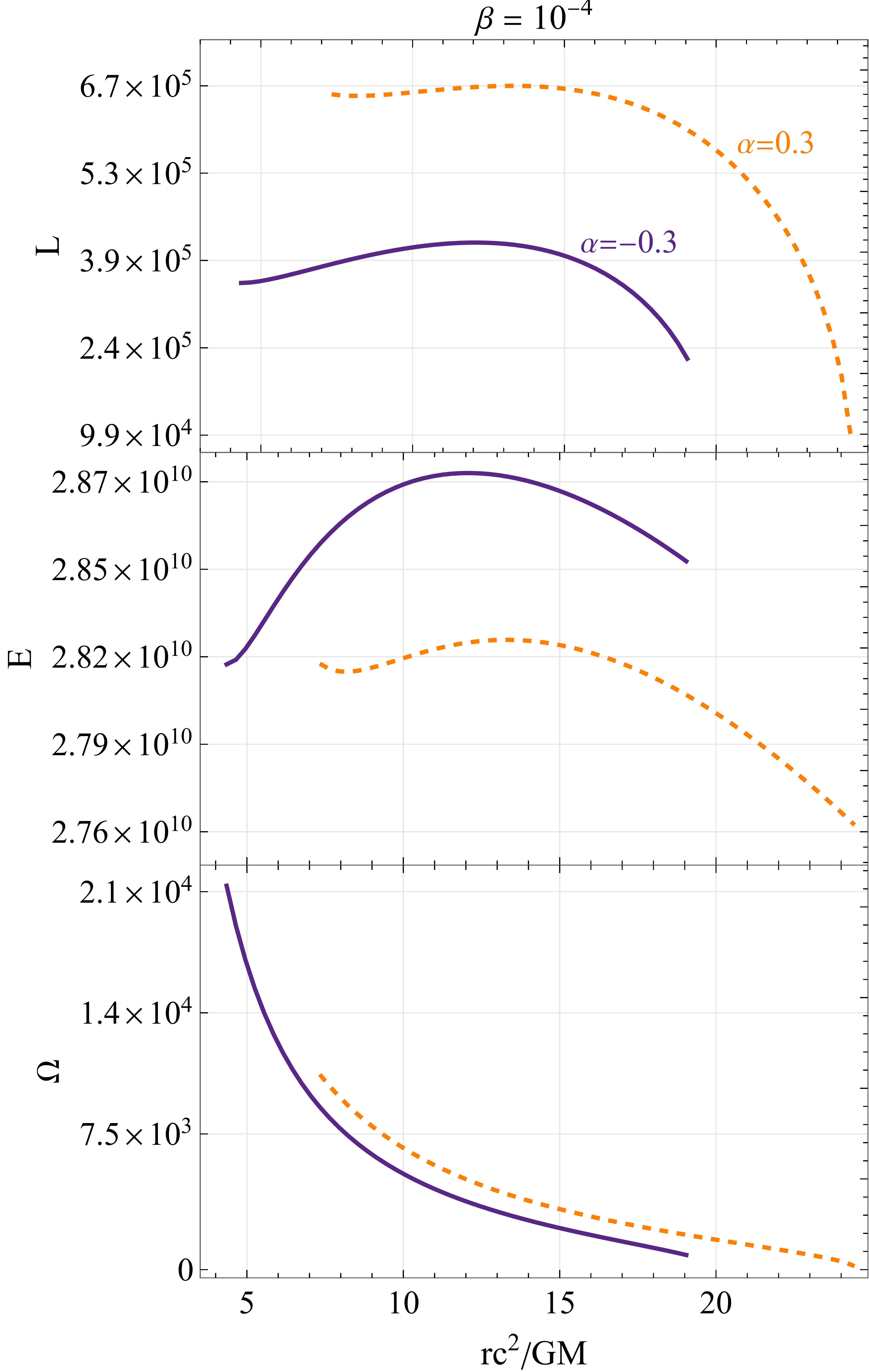}
    \caption{\label{fig:ELOmega_beta_pos} Equatorial ($y=0$) circular timelike invariants for positive quadrupolar distortion. From top to bottom the panels show the specific angular momentum $L$, the specific energy $E$, and the orbital angular velocity $\Omega=d\phi/dt$, plotted versus $rc^2/(GM)$. Left: $\beta=10^{-6}$; right: $\beta=10^{-4}$. Dashed orange and purple curves correspond to $\alpha=0.3$ and $\alpha=-0.3$, respectively. Curves are displayed only over the radial interval where the circular orbit expressions yield real and finite $E$, $L$, and $\Omega$; their common termination marks the outer limit of physically valid equatorial circular motion for the chosen parameters. For fixed $\beta$, larger $\alpha$ shifts $L$ and $\Omega$ to higher values and reduces $E$ (more strongly bound circular motion). Increasing $\beta$ strengthens the external quadrupolar distortion, leading to a more compact admissible interval and a more pronounced deformation of the orbital profiles. Curves are shown only where the circular orbit conditions \autoref{eq:ND} yield real and finite $E$, $L$, and $\Omega$ (equivalently, where the corresponding quantities remain positive).}

\end{figure*}

\begin{figure*}
    \centering 

     \includegraphics[width=0.45\textwidth]{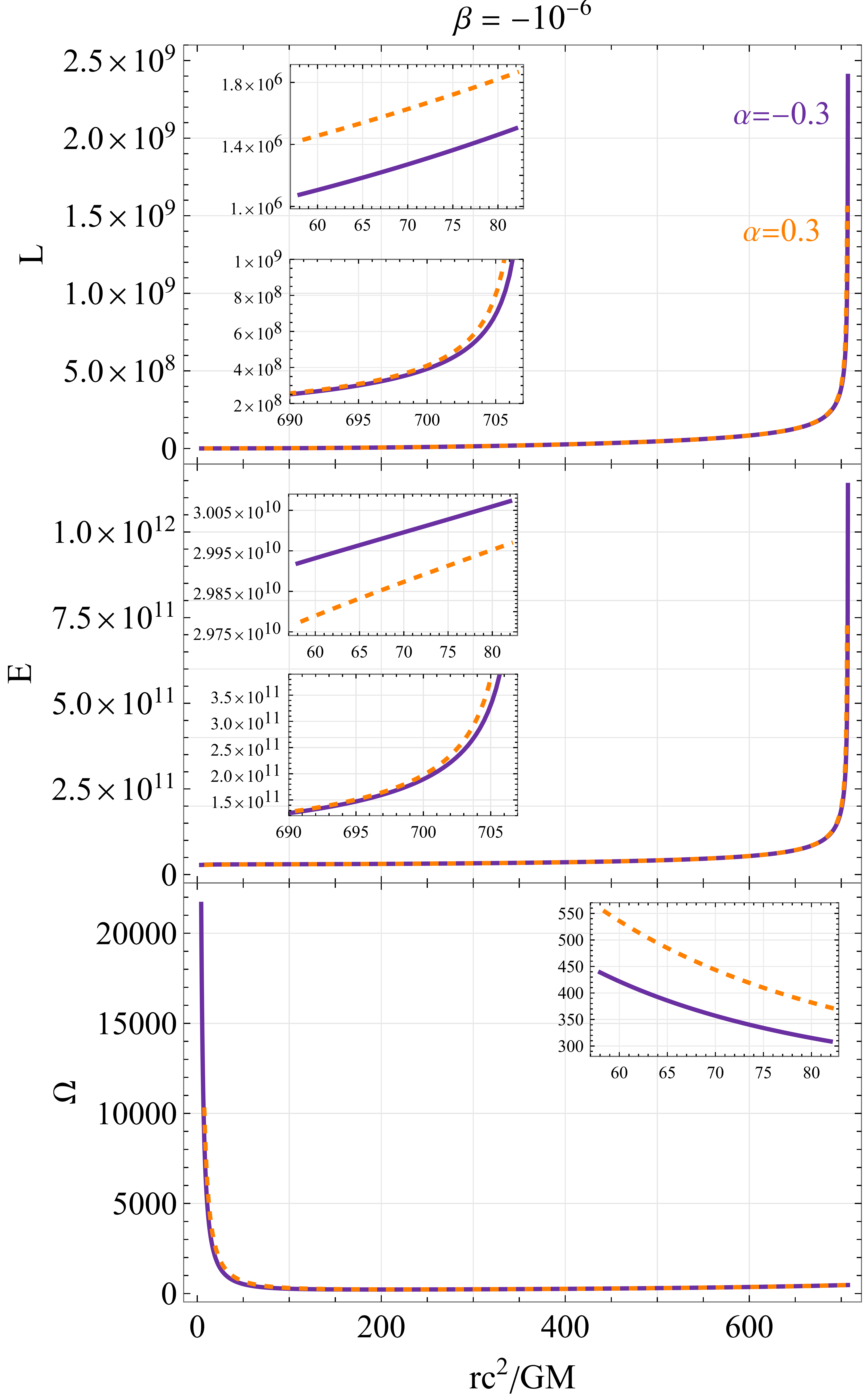}
     \includegraphics[width=0.445\textwidth]{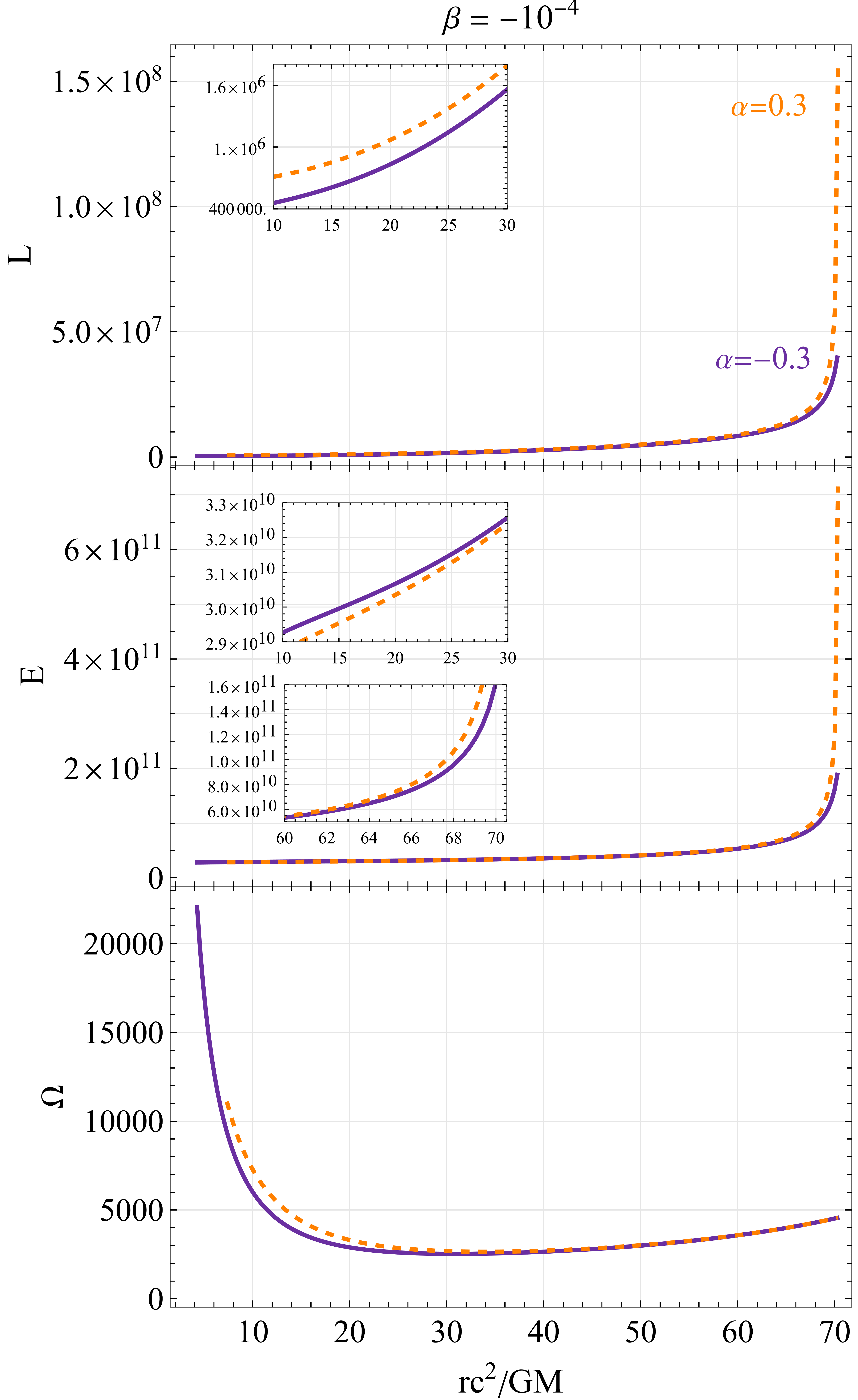}
    \caption{\label{fig:ELOmega_beta_neg}Equatorial ($y=0$) circular timelike-orbit invariants for negative quadrupolar distortion. Top row: specific angular momentum $L$; middle row: specific energy $E$; bottom row: angular velocity $\Omega=d\phi/dt$, all shown versus $rc^2/(GM)$. Left panel is for $\beta=-10^{-6}$; and right panel for $\beta=-10^{-4}$, implying effect of $\alpha$ is weaker as it is overwhelmed by the external field. Curves are plotted only where the circular orbit conditions are physically admissible, i.e. where the kinematic factors \autoref{eq:ND} satisfy $N_L>0$, $N_E>0$, and $D>0$ (with $x=r/M-1$). }
\end{figure*}

    

It is important to stress that this geometry is local in character. Since the distortion terms grow polynomially with $x$, the resulting spacetime is not intended as a globally asymptotically flat solution, but rather as an effective description of the neighborhood of a deformed compact source embedded in an external gravitational environment. In this sense, the solution may be viewed as a q-metric immersed in an external field, in the same spirit in which exact solutions have been used to model compact objects subjected to ambient fields \citep{1976JMP....17...54E}. Our aim is therefore to determine, for given values of $\alpha$ and $\beta$, the region in which this local quadrupolarly distorted geometry remains physically meaningful and self-consistent. 

\section{Behavior of physical quantities}\label{sec:behaviour}

A practical way to assess the physically meaningful domain of the local solution is to study the orbital observables associated with test particle motion in the distorted geometry. In particular, in this section for circular timelike geodesics we consider the conserved specific energy $E=-u_t$, the specific angular momentum $L=u_\phi$, and the angular velocity $\Omega=u^\phi/u^t$. For fixed values of the parameters $\alpha$ and $\beta$, the requirement that these quantities remain real and finite provides a first criterion for identifying the physically admissible region of the spacetime. A further diagnostic is provided by the radial behaviour of $\Omega$. In the seed configuration, namely in the absence of external distortion ($\beta=0$), $\Omega(x)$ is expected to decrease monotonically with $x$. By contrast, once the external quadrupolar field is switched on, $\Omega(x)$ may develop an extremum at some radius. We interpret the appearance of such a turning point as an indication that the influence of the external matter is no longer subdominant and that the local distorted solution is beginning to lose reliability. 

We now specialize to circular timelike motion in the equatorial plane, for which $y=0$ and $\dot y=0$. In the quadrupolar truncation, the distortion functions \autoref{eq:metricfunctions_xy} reduce to
\begin{align}\label{gammapsi1}
\hat{\psi} &= -\frac{\beta}{2}(x^2-1), \nonumber\\
\hat{\gamma} &= -2\beta(1+\alpha)x+\frac{\beta^2}{4}(x^2-1)^2.
\end{align}
Let
\begin{equation}\label{eq:Deltaa}
 a=1+\alpha, \qquad    \Delta=x^{2}-1,
\end{equation}
then for equatorial circular timelike orbits one obtains
\begin{align}\label{ELOmegaDef}
E^{2}&=\left(\frac{x-1}{x+1}\right)^a\frac{e^{-\beta\Delta}\,\Delta\,\bigl(x-a+\beta x\Delta\bigr)}{x-2a+2\beta x\Delta},\\
L^{2}&=M^{2}\Delta\left(\frac{x+1}{x-1}\right)^a\frac{e^{\beta\Delta}\,\bigl(a-\beta x\Delta\bigr)}{x-2a+2\beta x\Delta}\nonumber,
\end{align}
and 
\begin{equation}\label{eq:omega}
\Omega^{2}=\frac{1}{M^{2}(x+1)^{2}}\left(\frac{x-1}{x+1}\right)^{2a-1}
\,
\frac{e^{-2\beta\Delta}\bigl(a-\beta x\Delta\bigr)\bigl(x-a+\beta x\Delta\bigr)}{x-2a+2\beta x\Delta}.
\end{equation}
From these expressions, the valid circular orbit domain is controlled by three factors (besides trivial positive prefactors) as follows.
\begin{align}\label{eq:ND}
&N_L(x)=a-\beta x\Delta,\\
&N_E(x)=x-a+\beta x\Delta,\nonumber\\
&D(x)=x-2a+2\beta x\Delta,\nonumber
\end{align}
so that (up to strictly positive prefactors) $L^2\propto N_L/D$, $E^2\propto N_E/D$, and $\Omega^2\propto N_L/N_E$, and the basic existence requires
\begin{equation}
N_L>0,\qquad N_E>0,\qquad D>0.    
\end{equation}
For $\beta>0$ the factor $N_L$ necessarily loses positivity at a finite radius, implying an intrinsic outer kinematic cutoff for circular motion. This mechanism also explains the generic nonmonotonicity of $L(r)$: since $N_L\to0^+$ at the outer boundary, the angular momentum must turn over and decrease as the cutoff is approached, while $\Omega(r)$ remains positive and decreases throughout the admissible interval. More precisly, for $\beta>0$, $N_L(x)$ inevitably becomes negative at finite $x$, so the outer admissible edge is essentially set by
\begin{equation}
N_L(x_{\mathrm{out}})=0
\ \Rightarrow\ 
x_{\mathrm{out}}^{3}-x_{\mathrm{out}}=\frac{a}{\beta}
\ \Rightarrow\ 
x_{\mathrm{out}}\sim \left(\frac{1+\alpha}{\beta}\right)^{1/3}.
\end{equation}
As we can see, \autoref{fig:ELOmega_beta_pos} illustrates these features and shows that increasing $\beta$ compresses the admissible domain strongly, whereas increasing $\alpha$ (hence $a$) shifts the cutoff outward.

For negative quadrupolar distortion ($\beta<0$), we can write $N_L$ as $N_L=a+|\,\beta\,|x\Delta$ too see easier that $N_L$ is strictly positive for $x>1$, also $N_E$ stays positive throughout the relevant interval. Thus, the outer valid boundary is set by the timelike denominator $D$, which vanishes at a finite radius. As $D\to0^+$, both $E$ and $L$ diverge, while $\Omega$ remains finite because it does not contain $D$. This behavior is exactly what is seen in \autoref{fig:ELOmega_beta_neg}; in both panels, $E(r)$ and $L(r)$ increase sharply and blow up at the outer edge, whereas $\Omega(r)$ stays finite and, for sufficiently large $|\,\beta\,|$, can even develop a minimum within the admissible domain. The strong shrinking of the reliable interval when $|\,\beta\,|$ is increased follows from the large-$x$ balance in $D(x)=0$, which yields the scaling 
\begin{equation}
D(x_{\mathrm{out}})=0  \, \Rightarrow\ \, x_{\rm out}\,\sim(2\,|\,\beta\,|\,)^{-1/2},    
\end{equation}
(hence $r_{\rm out}\propto|\,\beta\,|^{-1/2}$). Finally, changing $\alpha$ mainly shifts the normalization and slightly shifts the cutoff radius through the $a$-dependence of $D(x)$; in particular, in the present cases the $\alpha=-0.3$ branch extends marginally farther in radius than $\alpha=0.3$.

To sum up, the sign of $\beta$ determines the mechanism of the outer cutoff: 
for fixed $|\beta|$, the meaningful equatorial circular orbit domain is generically much smaller for $\beta>0$ than for $\beta<0$: the outer cutoff is controlled by $N_L\to 0^{+}$ for $\beta>0$ ($x_{\mathrm{out}}\sim \beta^{-1/3}$) (suppressing $\Omega$ and forcing a turnover of $L$), whereas for $\beta<0$ it is controlled by $D\to 0^{+}$ at much larger radii ($x_{\mathrm{out}}\sim |\beta|^{-1/2}$) (driving $E$ and $L$ to diverge while $\Omega$ stays finite), making the $\beta<0$ case appear close to the undistorted ($\beta=0$) behavior in the strong field region.


\section{Stability of Equatorial Circular Orbits}\label{se:stability}

For timelike geodesics, the normalization condition is $u^\mu u_\mu=-1$. Using these conserved quantities, \autoref{ELOmegaDef}, the geodesic equation \citep{2022Univ....8..195F} can be recast as 
\begin{align}\label{radialgeneral}
M^2\left(\frac{x^2-1}{x^2-y^2}\right)^{\alpha(2+\alpha)} e^{2\hat{\gamma}} \dot{x}^2 + V^2(x,y;\dot y,L)=E^2,
\end{align}
where
\begin{align}\label{VtwoEI}
V^2 &=
\frac{x^2-1}{1-y^2}
\left(\frac{x^2-1}{x^2-y^2}\right)^{\alpha(2+\alpha)}
M^2 e^{2\hat{\gamma}} \dot{y}^2 \\
&\quad
+\left(\frac{x-1}{x+1}\right)^{2\alpha+1}
\frac{L^2 e^{4\hat{\psi}}}{M^2(1-y^2)(x+1)^2}
+\left(\frac{x-1}{x+1}\right)^{1+\alpha} e^{2\hat{\psi}}.\nonumber
\end{align}
The induced metric on the equatorial plane then takes the form
\begin{align}\label{EIeq}
ds^2=-A(x)\,dt^2+B(x)\,dx^2+C(x)\,d\phi^2,
\end{align}
with
\begin{align}
A(x) &= \left(\frac{x-1}{x+1}\right)^{1+\alpha} e^{2\hat{\psi}}, \nonumber\\
B(x) &= M^2\left(\frac{x+1}{x-1}\right)^{1+\alpha}
\left(\frac{x^2-1}{x^2}\right)^{\alpha(2+\alpha)}
e^{2\hat{\gamma}-2\hat{\psi}}, \nonumber\\
C(x) &= M^2(x^2-1)\left(\frac{x+1}{x-1}\right)^{1+\alpha} e^{-2\hat{\psi}}.
\end{align}
Accordingly, the radial equation becomes
\begin{align}\label{radialeq}
M^2\left(\frac{x^2-1}{x^2}\right)^{\alpha(2+\alpha)} e^{2\hat{\gamma}} \dot{x}^2 + V_{\rm Eff}(x;L)=E^2,
\end{align}
where the effective potential for timelike motion is
\begin{align}\label{Vei}
V_{\rm Eff}(x;L)
=
\left(\frac{x-1}{x+1}\right)^{1+\alpha} e^{2\hat{\psi}}
\left[
1+
\frac{L^2 e^{2\hat{\psi}}}{M^2(x+1)^2}
\left(\frac{x-1}{x+1}\right)^\alpha
\right].
\end{align}
To make the marginal stability condition explicit, then by using \autoref{eq:Deltaa}, then on the equatorial plane, the metric functions entering the effective potential can be written as
\begin{align}
A(x)&:=\left(\frac{x-1}{x+1}\right)^a e^{-\beta\Delta},\\
C(x)&:=M^2\Delta\left(\frac{x+1}{x-1}\right)^a e^{\beta\Delta}.
\end{align}
Then the effective potential for timelike equatorial motion takes the standard form
\begin{align}
V_{\rm Eff}(x;L)=A(x)\left(1+\frac{L^2}{C(x)}\right).
\end{align}
A circular orbit at $x=x_0$ is determined by
\begin{align}
V_{\rm Eff}(x_0;L)=E^2,\qquad
\frac{dV_{\rm Eff}}{dx}(x_0;L)=0.
\end{align}
Introducing the logarithmic derivatives
\begin{align}
p(x):=\frac{A'(x)}{A(x)},\qquad
q(x):=\frac{C'(x)}{C(x)},
\end{align}
the second condition yields
\begin{align}
L^2=\frac{C\,p}{q-p},\qquad
E^2=\frac{A\,q}{q-p}.
\end{align}
For the present geometry one finds
\begin{align}
p(x)&=\frac{2\big(a-\beta x\Delta\big)}{\Delta},\\
q(x)&=\frac{2\big(x-a+\beta x\Delta\big)}{\Delta},
\end{align}
Marginal stability is determined by the additional condition
\begin{align}
\frac{d^2V_{\rm Eff}}{dx^2}=0.
\end{align}
After substituting the circular orbit expressions above, this condition can be written exactly as
\begin{align}\label{Vppcompact}
\frac{d^2V_{\rm Eff}}{dx^2}
=
\frac{A}{q-p}\Big[q\,p'-p\,q'+pq(q-p)\Big].
\end{align}
Hence the marginally stable radii satisfy
\begin{align}\label{mscompact}
q\,p'-p\,q'+pq(q-p)=0,
\end{align}
where differentiating of $L$ given by \autoref{ELOmegaDef} gives
\begin{align}
\frac{dL^2}{dx}
=
\frac{C}{(q-p)^2}\Big[q\,p'-p\,q'+pq(q-p)\Big].
\end{align}
Therefore, on the physical branch of circular timelike orbits, the two criteria
\begin{align}
\frac{d^2V_{\rm Eff}}{dx^2}=0,
\qquad\Longleftrightarrow\qquad
\frac{dL^2}{dx}=0,
\end{align}
are exactly equivalent. This explains why the extrema of $L$ reproduce the same marginally stable radii as the effective-potential analysis which used in \citep{2022Univ....8..195F}.

Substituting the explicit forms of $p$ and $q$, \autoref{mscompact} becomes
\begin{align}\label{mslog}
2\beta x+\frac{a+1}{x+1}-\frac{a-1}{x-1}
-\frac{\beta(3x^2-1)}{a-\beta x\Delta}
-\frac{1+2\beta(3x^2-1)}{x-2a+2\beta x\Delta}=0.
\end{align}
Equivalently, after clearing denominators, one obtains the algebraic equation
\begin{align}\label{mspoly}
P(x;a,\beta)&:=
4\beta^3x^3\Delta^3
+6\beta^2x^2\Delta^2(x-2a)\nonumber\\
&\quad +4\beta x\Delta(x^2-3ax+3a^2)\nonumber\\
&\quad -a(x^2-6ax+4a^2+1)=0.
\end{align}
This equation is cubic in $\beta$ and ninth order in $x$, which is why the marginally stable radii are, in general, obtained numerically.

Nevertheless, in the undistorted limit $\beta=0$, \autoref{mspoly} reduces to
\begin{align}
x^2-6ax+4a^2+1=0,
\end{align}
so that the two marginally stable radii are obtained analytically as
\begin{align}\label{msbeta0}
x_{\rm ISCO}^{(0)\pm}=3a\pm\sqrt{5a^2-1}.
\end{align}
Thus, whenever $5a^2>1$, the two branches are real. In particular, for $\alpha=0$ one recovers
\begin{align}
x_{\rm ISCO}^{(0)-}=1,\qquad x_{\rm ISCO}^{(0)+}=5,
\end{align}
and only the outer branch corresponds to the Schwarzschild ISCO.

\begin{figure}
    \centering 
    \includegraphics[width=0.51\textwidth]{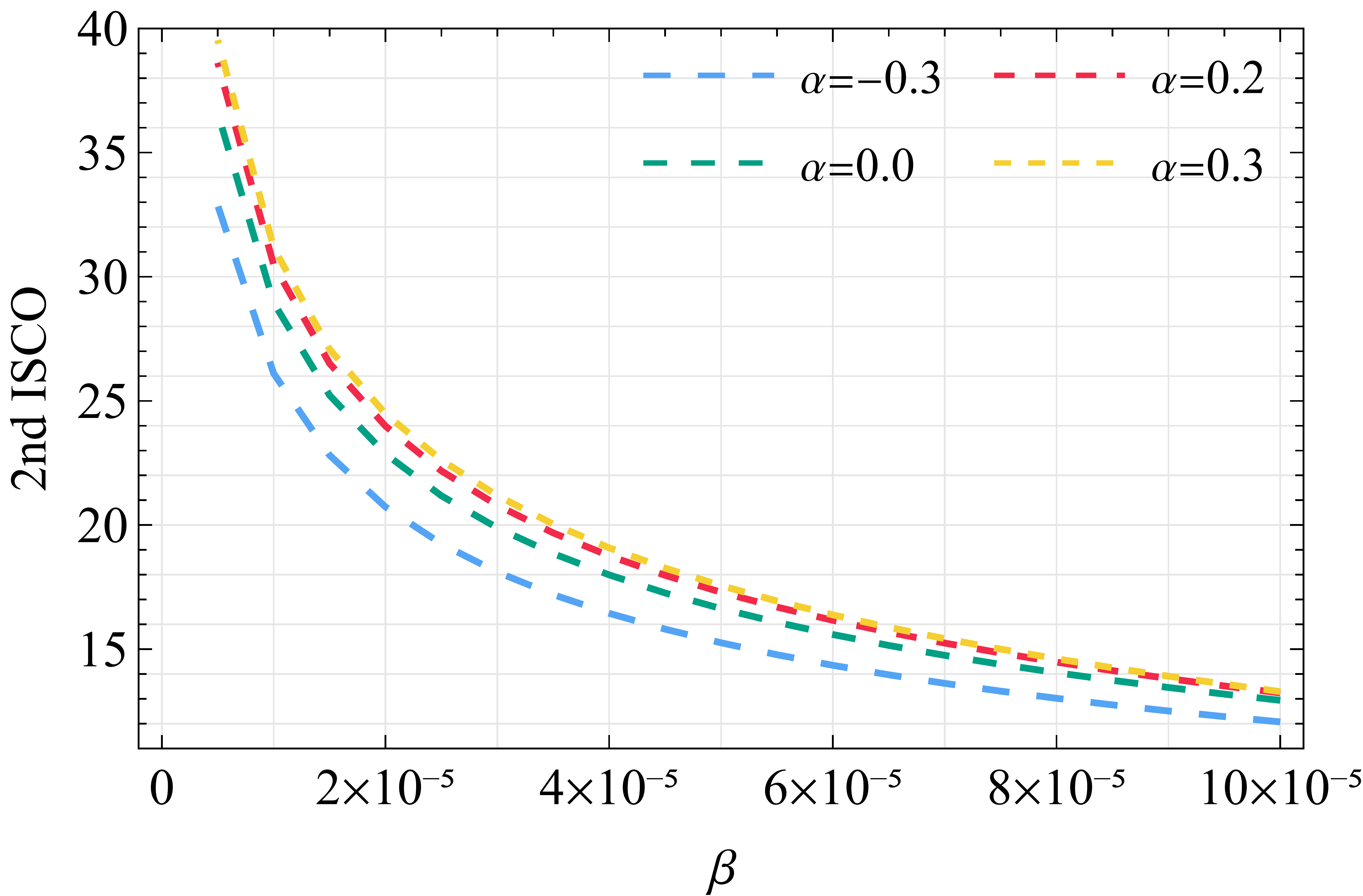}
    \includegraphics[width=0.51\textwidth]{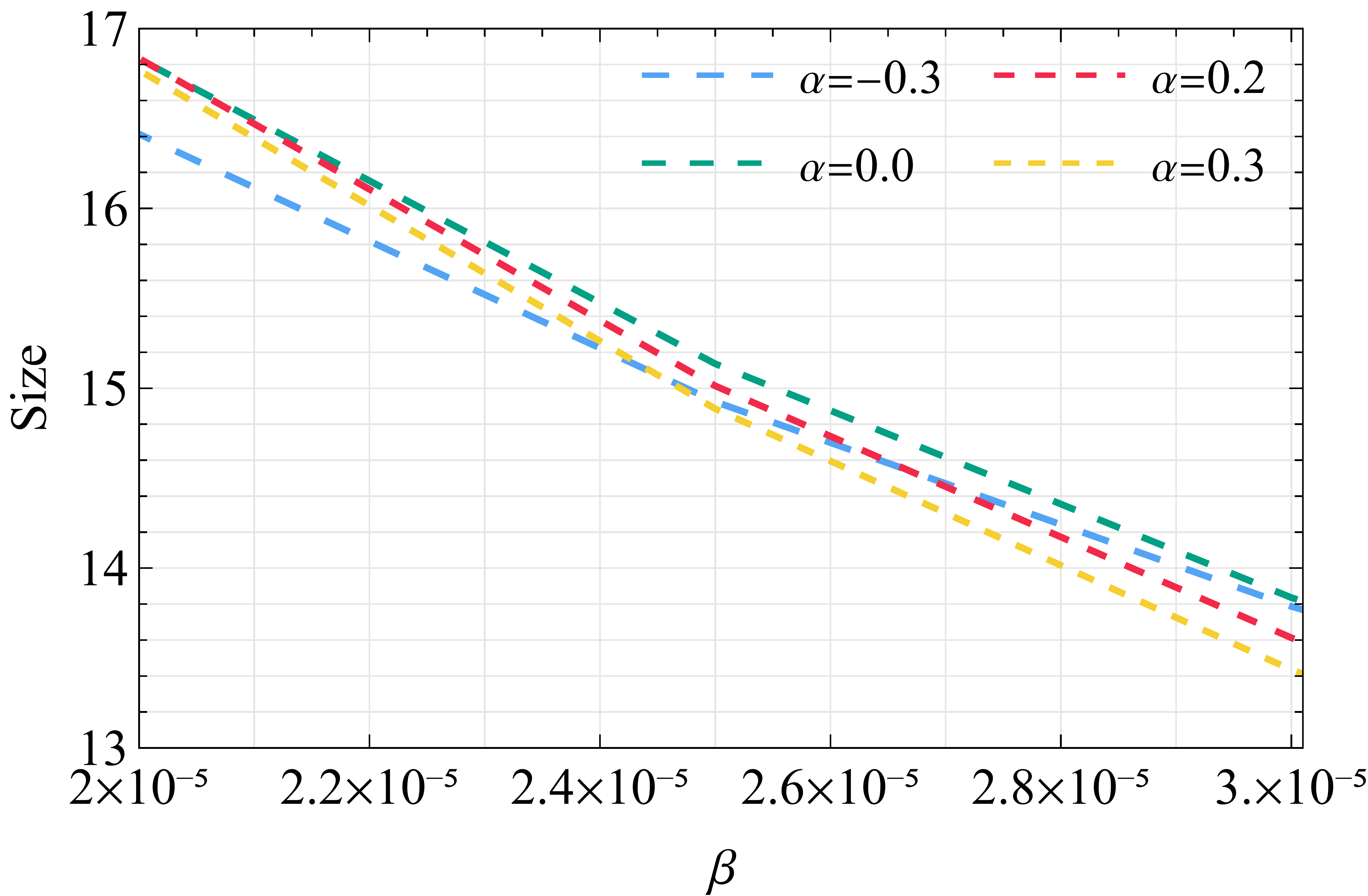}
  \caption{\label{fig:secondISCO_betafig1}Upper panel: second (outer) marginally stable radius $x_{\rm ISCO}^{(2)}$ as a function of the positive quadrupolar distortion parameter $\beta$ for selected values of the deformation parameter $\alpha$. The outer branch moves inward monotonically as $\beta$ increases and, for fixed $\beta$, is located at larger radius for larger $\alpha$. Lower panel: radial size $\Delta x=x_{\rm ISCO}^{(2)}-x_{\rm ISCO}^{(1)}$ of the corresponding annular stable region. The width decreases with increasing $\beta$, showing that the external quadrupolar field progressively narrows the ring-like domain of stable circular motion.}
\end{figure}

Moreover, since the roots in \autoref{msbeta0} are simple provided $5a^2\neq 1$, the implicit-function theorem guarantees that, for sufficiently small $|\,\beta\,|$, both branches deform smoothly as
\begin{align}
x_{\rm ISCO}^{\pm}(\beta)
=
x_{\rm ISCO}^{(0)\pm}
+\beta\,x_1^{\pm}
+\mathcal{O}(\beta^2),
\end{align}
with
\begin{align}
x_1^{\pm}
=
-\left.
\frac{\partial_\beta P(x;a,\beta)}{\partial_x P(x;a,\beta)}
\right|_{(x,\beta)=(x_{\rm ISCO}^{(0)\pm},0)}.
\end{align}
A direct evaluation yields
\begin{align}
x_1^{\pm}
=
\frac{2x_{\rm ISCO}^{(0)\pm}\Big[\big(x_{\rm ISCO}^{(0)\pm}\big)^2-1\Big]
\Big[3a^2-3a\,x_{\rm ISCO}^{(0)\pm}+\big(x_{\rm ISCO}^{(0)\pm}\big)^2\Big]}
{a\Big(x_{\rm ISCO}^{(0)\pm}-3a\Big)}.
\end{align}
Hence the two marginally stable radii admit an explicit perturbative expansion in the external quadrupolar distortion, even though the full problem is not solvable in closed form. In the rest of this section, we analyze equatorial circular timelike motion and the location of the marginally stable radii in the quadrupolarly distorted spacetime. For positive quadrupolar distortion, the marginal stability equation admits an additional solution at large radius. In the weak-distortion regime, this second marginally stable radius can be obtained analytically from the large-$x$ limit of the stability condition, yielding
\begin{equation}
x_{\rm ISCO}^{(2)}=
\left(\frac{1+\alpha}{4\beta}\right)^{1/3}
-\frac{9}{8}(1+\alpha)
+{\cal O}(\beta^{1/3}).
\end{equation}
This shows that $x_{\rm ISCO}^{(2)}$ diverges as $\beta\to0^+$ and therefore disappears in the undistorted limit. Hence, the outer marginally stable branch is generated by the external quadrupolar field. The numerical results displayed in \autoref{fig:secondISCO_betafig1} are fully consistent with this behavior: for all values of $\alpha$ considered, the second marginally stable radius decreases monotonically as $\beta$ increases, while at fixed $\beta$ it shifts outward for larger $\alpha$.

If the interval bounded by the relevant inner and outer marginally stable radii is interpreted as a ring-like accretion region, then its radial extent
\begin{equation}\label{eq:size}
\Delta x = x_{\rm ISCO}^{(2)} - x_{\rm ISCO}^{(1)}, 
\end{equation}
also decreases monotonically with $\beta$, as shown in the lower panel of \autoref{fig:secondISCO_betafig1}. This reflects the fact that the outer stability boundary is considerably more sensitive to the external distortion than the inner one as expected. As a consequence, increasing $\beta$ compresses the admissible annular region and pushes it closer to the compact object. When an additional marginally stable root appears close to $x=1$, the quantity $\Delta x$ is defined using the larger inner boundary that delimits the physically relevant outer stable zone.

\begin{figure}
    \centering 
    \includegraphics[width=0.49\textwidth]{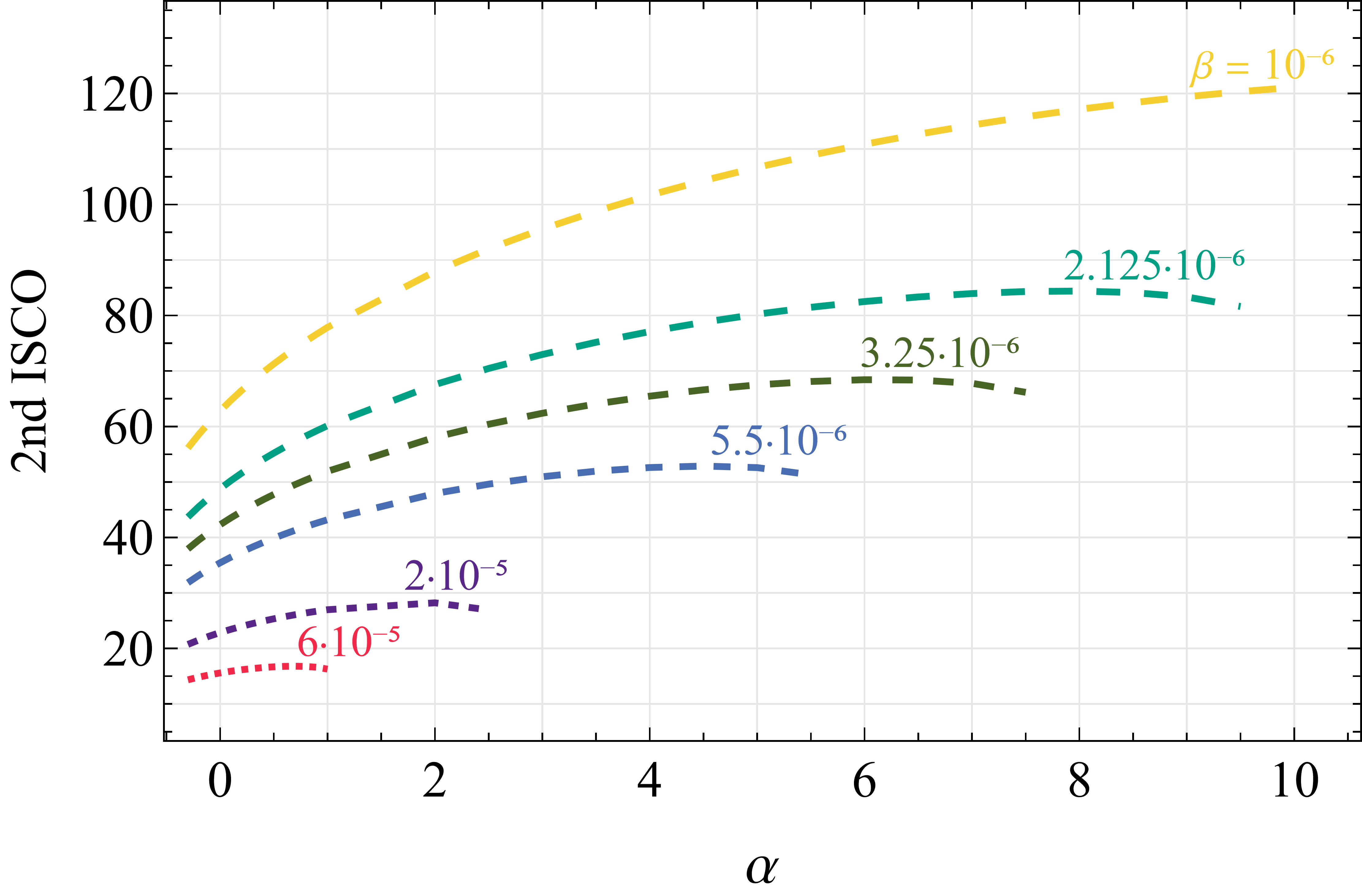}
    \includegraphics[width=0.49\textwidth]{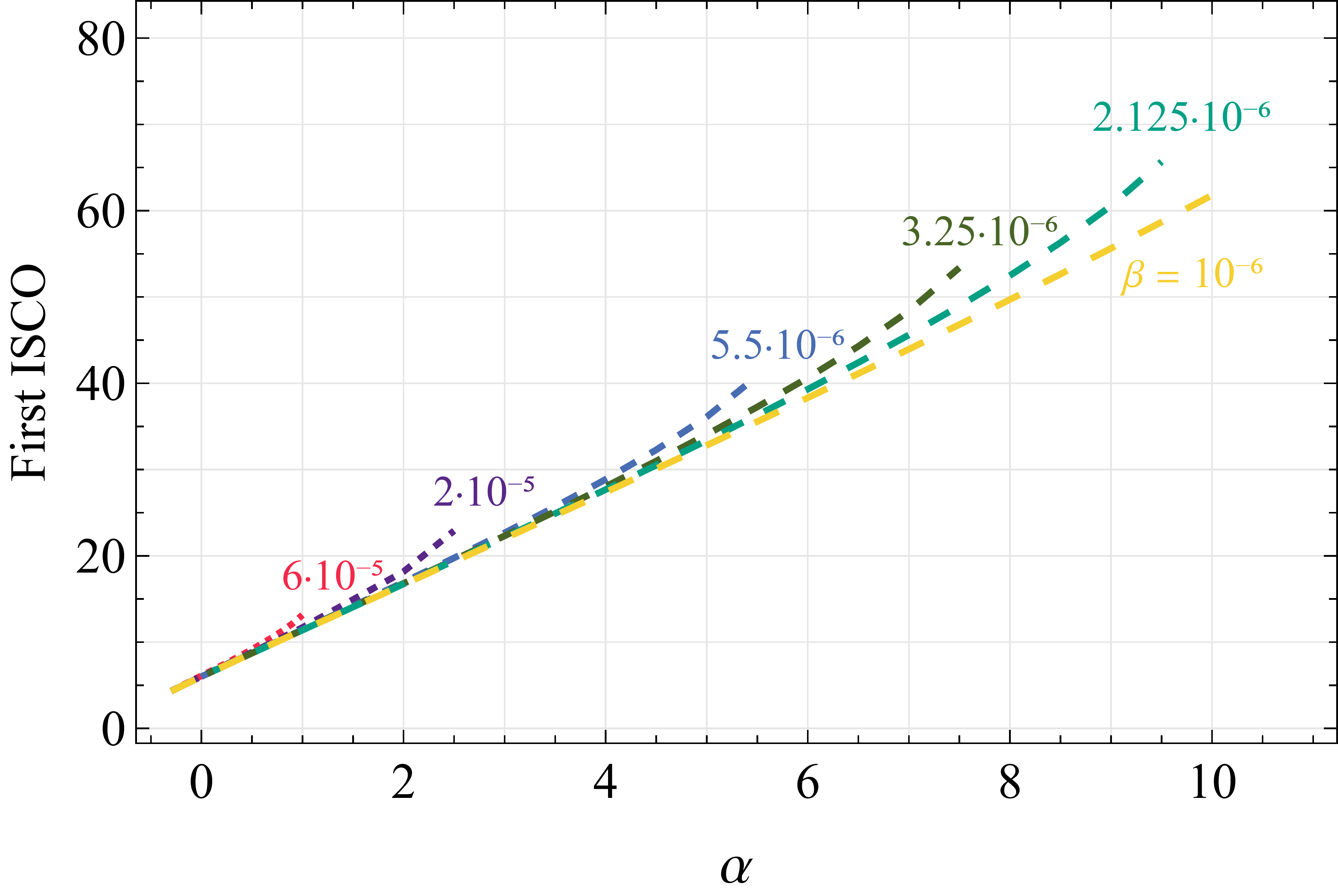}
    \includegraphics[width=0.49\textwidth]{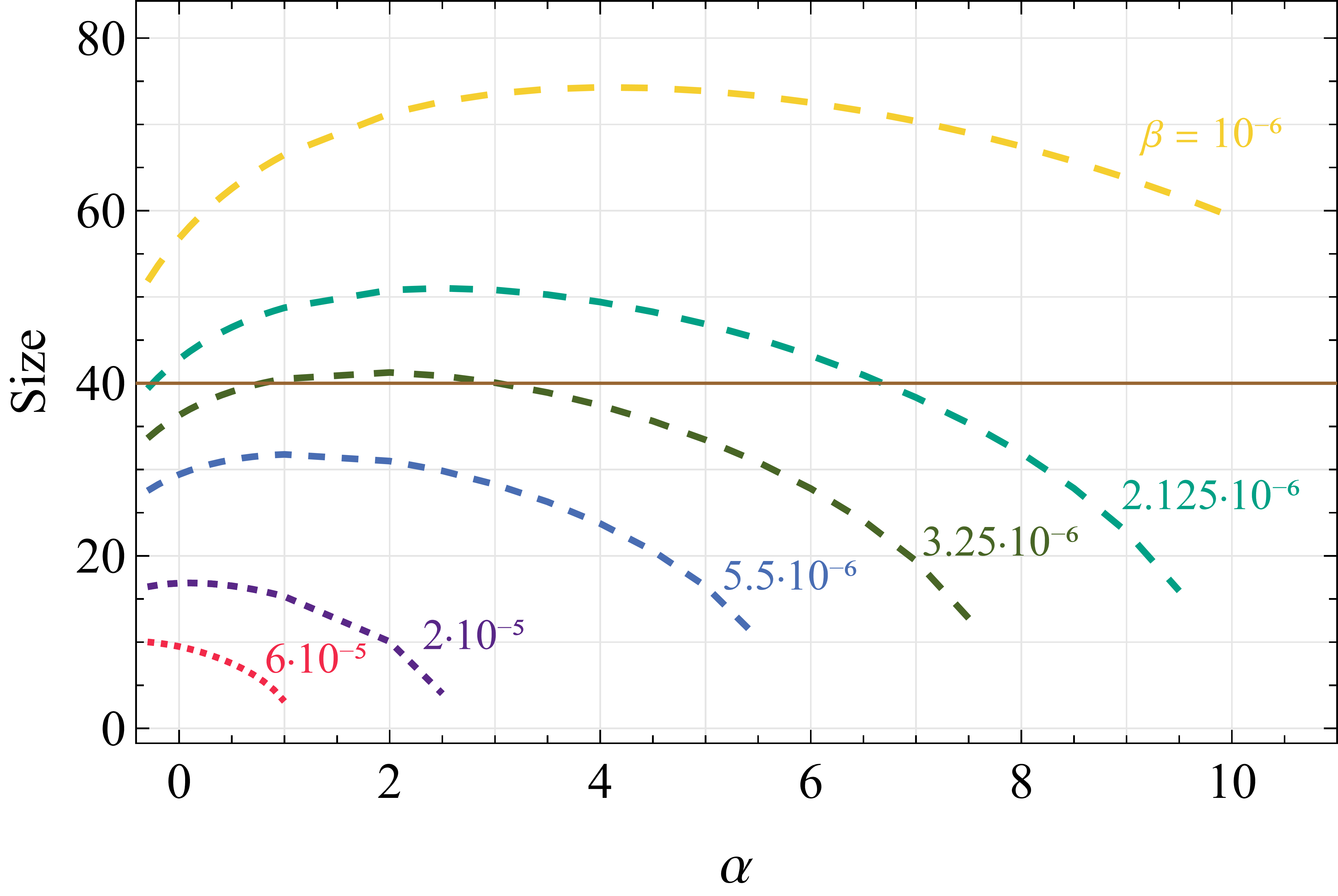}
    \caption{\label{fig:alphaISCO_alpha}Marginally stable radii as functions of the deformation parameter $\alpha$ for selected positive values of the external quadrupolar distortion $\beta$. Top panel: outer marginally stable radius $x_{\rm ISCO}^{(2)}$. Middle panel: inner boundary $x_{\rm ISCO}^{(1)}$ of the physically relevant outer stable region. Bottom panel: radial size $\Delta x=x_{\rm ISCO}^{(2)}-x_{\rm ISCO}^{(1)}$ of the corresponding annular stable domain. For each fixed $\beta$, the two outer marginally stable radii approach each other and merge at a critical value of $\alpha$, beyond which the annular region ceases to exist. Smaller values of $\beta$ shift the outer branch to larger radii and allow the annular stable region to persist over a wider range of $\alpha$.}
\end{figure}

Figure~\ref{fig:alphaISCO_alpha} shows the dependence of the marginally stable radii on the intrinsic deformation parameter $\alpha$ for several fixed positive values of the external quadrupolar distortion $\beta$. For $\beta>0$, the marginal stability equation may admit three exterior roots. In the present discussion, $x_{\rm ISCO}^{(1)}$ denotes the inner boundary of the physically relevant outer stable region, while $x_{\rm ISCO}^{(2)}$ denotes its outer boundary. The upper panel shows that the second marginally stable radius moves outward as $\alpha$ increases, and is located at larger radii for smaller values of $\beta$. This behavior is consistent with the large-radius asymptotic scaling
\begin{equation}
x_{\rm ISCO}^{(2)} \sim \left(\frac{1+\alpha}{4\beta}\right)^{1/3},
\end{equation}
which captures the leading effect of the external quadrupolar distortion. The middle panel shows that $x_{\rm ISCO}^{(1)}$ also increases with $\alpha$, as expected from its continuous deformation from the undistorted marginally stable branch.

A particularly important feature is that, for each fixed $\beta$, the two outer marginally stable radii approach one another and eventually coalesce at a critical value $\alpha=\alpha_c(\beta)$. This occurs when the marginal stability polynomial develops a double root, namely when
\begin{equation}
P(x;1+\alpha,\beta)=0,
\qquad
\partial_x P(x;1+\alpha,\beta)=0.    
\end{equation}
At this point, the annular stable region disappears. This is reflected in the lower panel, where the radial size \autoref{eq:size} decreases to zero at the endpoint of each curve. Therefore, the figure shows that the existence and extent of the ring-like stable region are controlled jointly by the intrinsic deformation $\alpha$ and the external quadrupolar distortion $\beta$.

\section{Accretion diagnostics of the validity domain}\label{sec:diskdomain}

The geodesic analysis above suggests that the distorted spacetime admits a physically meaningful description only within a finite radial domain as it expected from its construction. A more stringent and astrophysically relevant test of this local validity is provided by thin accretion configurations, whose structure is governed by the properties of equatorial circular motion. We therefore examine whether the same finite domain is selected by the dynamical and radiative quantities entering the thin disk description or can be modified further. 
Therefore, we adopt the thin disk formalism, but we apply it only in the region where (i) the local metric is meaningful, (ii) circular geodesics are admissible,
and (iii) the orbits are stable.

\subsection{Radiatively efficient thin disk}\label{sec:disk}

The accretion flow as a geometrically thin, radiatively efficient disk embedded in a stationary, axisymmetric spacetime. Stationarity and axisymmetry background space-time provides two Killing vectors, $\xi^\mu_{(t)}$ and $\xi^\mu_{(\phi)}$, which in turn define conserved energy and angular momentum currents associated with the stress energy tensor. The thin disk approximation restricts the dynamics to a narrow neighborhood of the equatorial plane ($H/r\ll1$) and enforces quasi-circular motion with a slow radial drift. In this setting, the disk structure can be formulated as a balance between (i) conserved radial transport of rest mass, energy, and angular momentum, and (ii) local radiative losses from the disk surfaces.

 The Page-Thorne relation then follows by combining the conserved energy and angular momentum fluxes with the inner boundary condition at $r_0=r_{\rm ISCO}$, yielding the radiative flux profile $F(r)$ emitted from each face of the disk.

\begin{figure*}
    \centering
    \includegraphics[width=0.45\linewidth]{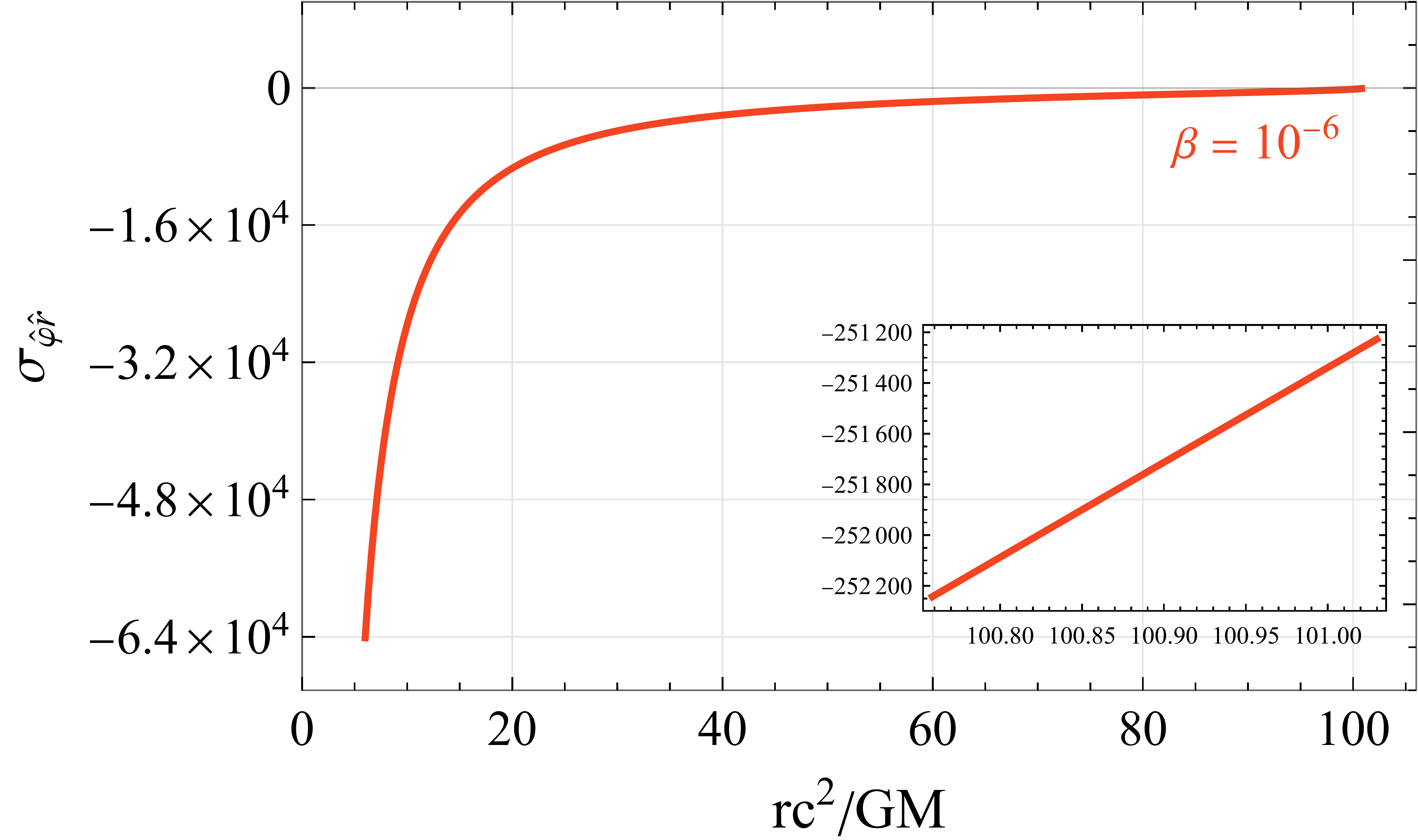}
    \includegraphics[width=0.45\linewidth]{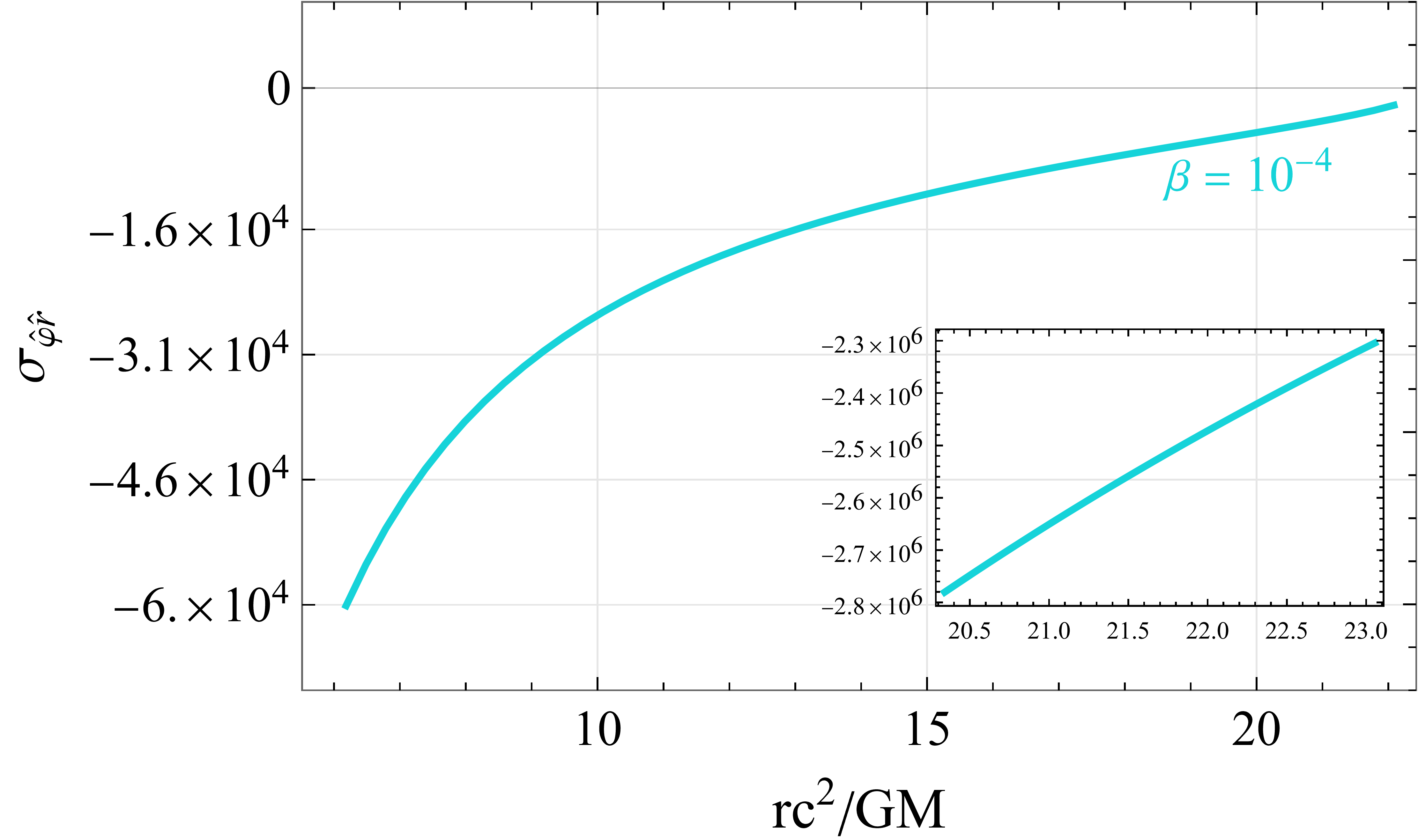}\\
    \includegraphics[width=0.45\linewidth]{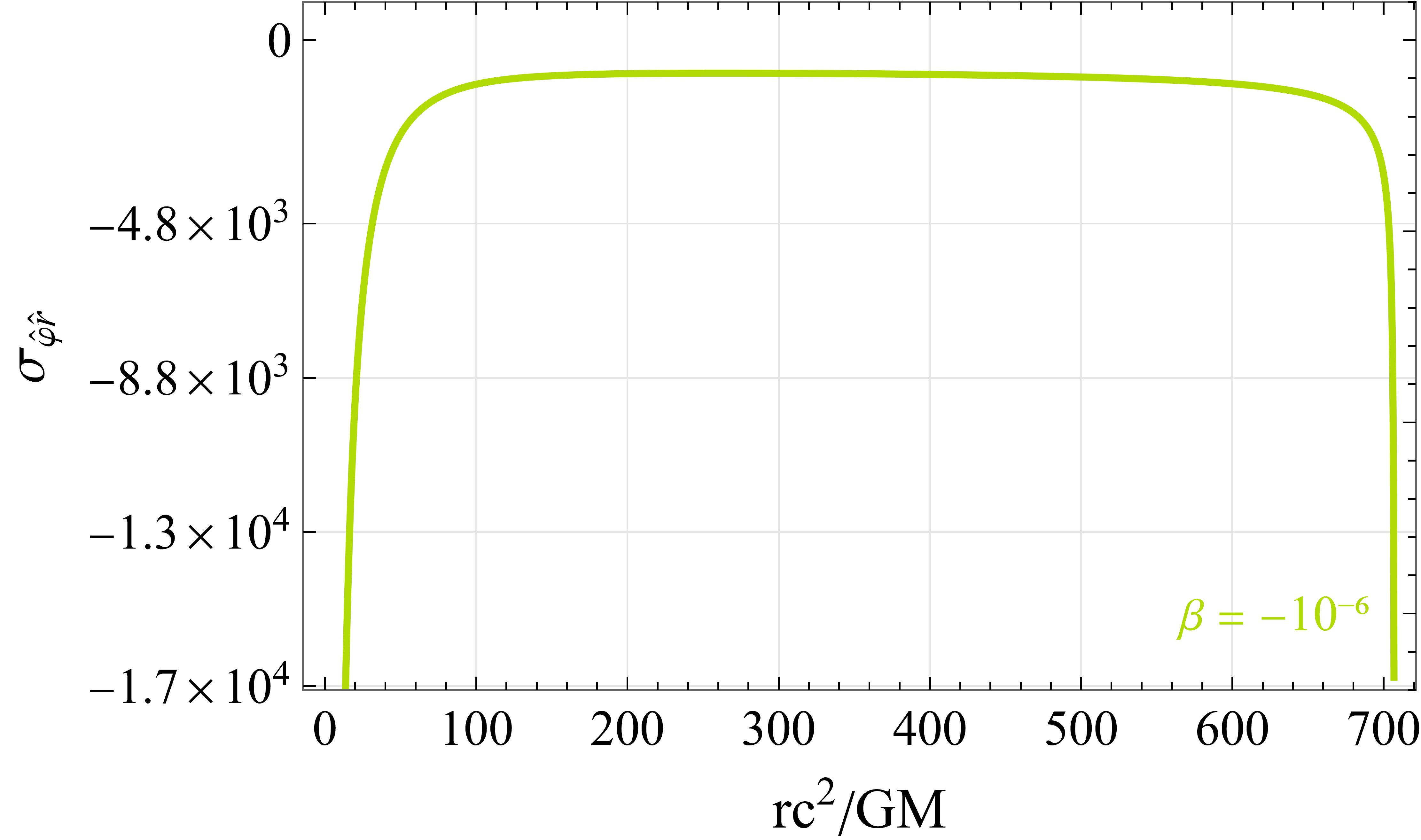}
    \includegraphics[width=0.45\linewidth]{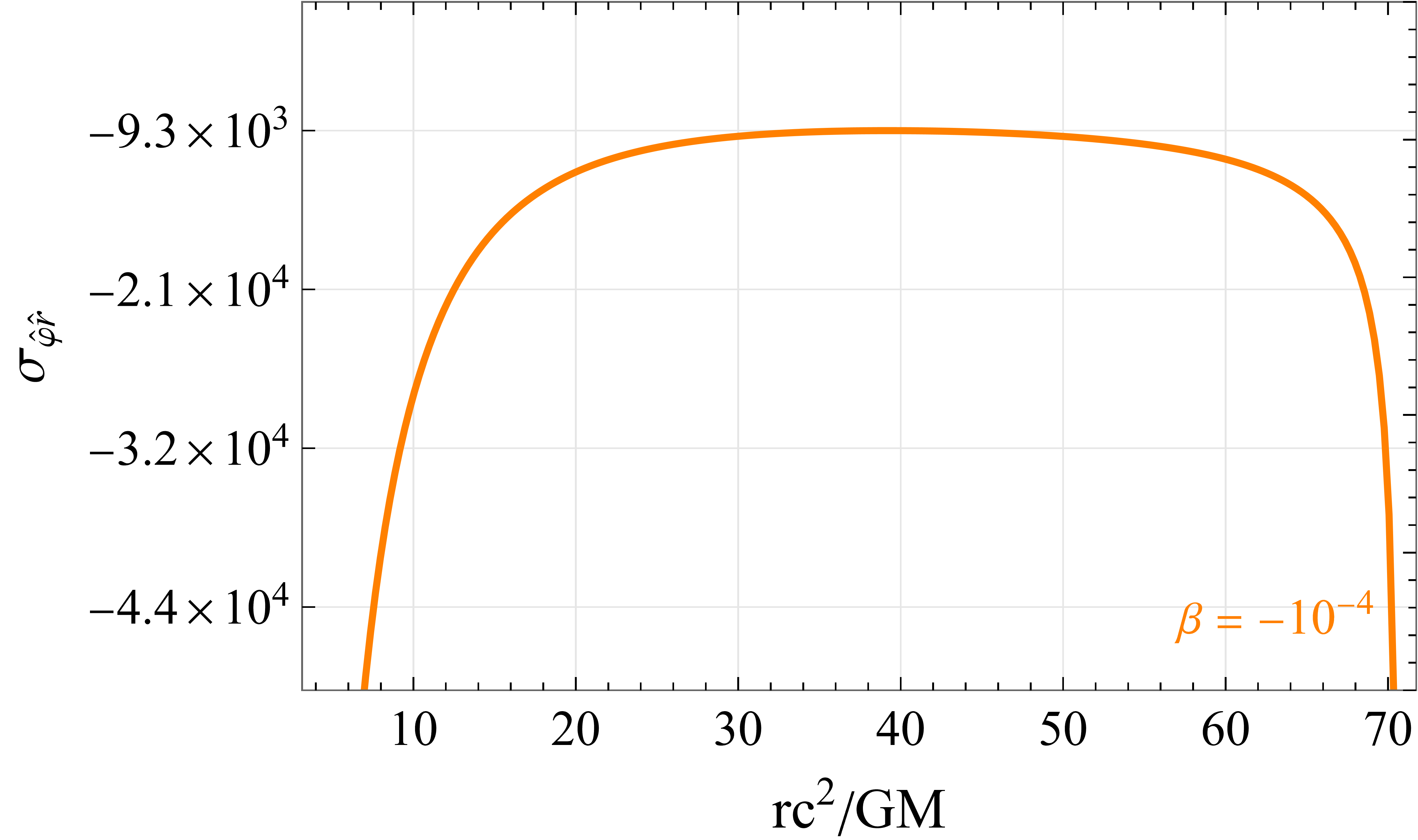}
    \caption{\label{fig:shear}Geometry based diagnostic panel. Local orthonormal shear component $\sigma_{\hat r\hat\phi}$ as a function of $rc^2/(GM)$ for representative values of the distortion parameters for $\alpha=0$. In all cases, the shear remains negative throughout the physically admissible region. For some parameter choices, $\sigma_{\hat r\hat\phi}$ tends to zero at a finite radius (highlighted in the insets), signaling the loss of differential rotation and providing an outer bound for the local thin disc solution. For other parameter choices, the admissible radial interval ends before the zero-shear point is reached, leading instead to a sharp edge behavior near the outer boundary.}
    \end{figure*}

In this model, the shear stress is supposed to be a form of viscosity, which is responsible for transporting angular momentum and energy outward and accreting matter inward. Also, it heats up the gas locally, introduces viscosity through a so-called $\alpha$-prescription. In an effort to modify the $\alpha$-prescription, in the literature in general, there are two ways: by considering $\alpha$ as a function of radius or keeping $\alpha$ a constant and multiplying this by a factor. However, global magnetohydrodynamic simulations argued that $\alpha$ is a function of $r$ (e.g., \cite{1996MNRAS.281L..21A,f4a0f084261d4399b3ed571587d7f17c,10.1093/mnras/sts185}).

The radial structure of a stationary, axially symmetric thin disc is governed by rest-mass conservation together with the energy and momentum equations obtained from the conservation of the stress-energy tensor. The conservation of particle number is
\begin{equation}\label{restmasscon}
(\rho u^{\mu})_{;\mu}=0,
\end{equation}
where $u^{\mu}$ is the four-velocity of the fluid and $\rho$ is the rest-mass density. The projection of $T^{\mu\nu}{}_{;\nu}=0$ along the four-velocity,
\begin{equation}\label{energycon}
u_{\mu}T^{\mu\nu}{}_{;\nu}=0,
\end{equation}
yields the local energy equation, whereas the projection orthogonal to the flow,
\begin{equation}\label{NSE}
h_{\mu\sigma}(T^{\sigma\nu})_{;\nu}=0,
\end{equation}
gives the relativistic Euler-Navier-Stokes equation. Here
\begin{equation}
h^{\mu\nu}=g^{\mu\nu}+u^{\mu}u^{\nu}
\end{equation}
is the projection tensor onto the instantaneous rest space of the fluid. The stress-energy tensor is decomposed as
\begin{align}
T^{\mu\nu}
=\varepsilon\,u^{\mu}u^{\nu}
+p\,g^{\mu\nu}+q^{\mu}u^{\nu}+q^{\nu}u^{\mu}+S^{\mu\nu},
\end{align}
where $\varepsilon=\varepsilon_0+p$ is the enthalpy density, $\varepsilon_0$ is the total energy density measured in the fluid frame, $p$ is the pressure, $q^{\mu}$ is the heat flux orthogonal to $u^{\mu}$, and $S^{\mu\nu}$ is the viscous stress tensor.

\subsection{Behavior of the orthonormal shear component}
In the absence of bulk viscosity is given by
\begin{equation}
S^{\mu\nu}=-2\lambda\,\sigma^{\mu\nu},
\end{equation}
where $\lambda$ is the dynamical viscosity and $\sigma^{\mu\nu}$ is the shear tensor.

Under the standard assumptions of the thin disc model, the dynamically relevant viscous component governing angular momentum transport is the $r\phi$ component,
\begin{align}\label{shearsigmahat}
\sigma_{r\phi}
=
\frac{1}{2}
\left(
u_{r;\beta}h^{\beta}{}_{\phi}
+
u_{\phi;\beta}h^{\beta}{}_{r}
\right)
-
\frac{1}{3}
h_{r\phi}\,u^{\beta}{}_{;\beta}.
\end{align}
Since coordinate components depend on the chosen basis, the physically meaningful quantity is the shear measured in the local orthonormal frame comoving with the fluid. We therefore consider
\begin{align}\label{eq:sigmahat}
\sigma_{\hat r\hat\phi}
=e^{r}{}_{\hat r}\,
e^{\phi}{}_{\hat\phi}\,
\sigma_{r\phi},
\end{align}
where $e^{\mu}{}_{\hat\mu}$ denotes the orthonormal tetrad associated with the metric. For the diagonal metric considered here, this transformation simply rescales the coordinate component by the appropriate metric factors. However, in general, one can obtain the nonholonomic basis from the holonomic one by applying the Gram-Schmidt process. In the slow drift approximation, where terms quadratic in $u^{r}$ are neglected, the corresponding expression reduces in the Kerr limit to the familiar result given in \cite[equation (5.4.6)]{1973blho.conf..343N}. Therefore, this $\sigma_{\hat r\hat\phi}$ is another diagnostic of the physically meaningful radial domain. Since this quantity measures the local differential rotation of the flow, its behavior is directly tied to the radial variation of the angular velocity. In the asymptotically flat geometries, like Schwarzschild and the undistorted q-metric ($\beta=0$), the local orthonormal shear component decreases with radius and tends to zero only asymptotically. Thus, in those cases the shear does not single out any finite outer cutoff. A qualitatively different behavior emerges once the external quadrupolar distortion is switched on: $\sigma_{\hat r\hat\phi}$ may vanish at a finite radius. Since the seed q-metric remains asymptotically flat for all admissible values of $\alpha$, and varying $\alpha$ changes only the location of the profile without modifying its qualitative pattern, this finite radius effect should be attributed to the external distortion parameter $\beta$. For this reason, in \autoref{fig:shear} we set $\alpha=0$ in order to isolate the role of $\beta$ as clearly as possible.

The quantity $\sigma_{\hat r\hat\phi}$ is directly tied to the radial variation of the angular velocity and remains negative throughout the physically admissible disc region, as expected for a differentially rotating flow. When $\sigma_{\hat r\hat\phi}$ approaches zero at a finite radius, the local shear of the flow disappears there, consistently with the appearance of an extremum in $\Omega$, \autoref{eq:omega}. We therefore interpret the corresponding radius as an independent indicator of the outer boundary beyond which the local distorted spacetime no longer provides a reliable thin disc description. For parameter choices in which the admissible disc region terminates before the zero-shear point is reached, $\sigma_{\hat r\hat\phi}$ remains negative but develops a sharp edge behavior near the boundary. In both situations, the shear profile supports the conclusion that the distorted geometry is physically meaningful only within a finite radial interval.



\subsection{Relativistic thin disk structure equations}\label{eq:equations}

Following the relativistic thin disk formalism of \citet{1973blho.conf..343N,1974ApJ...191..499P}, vertically average the conservation laws Equations \eqref{restmasscon}-\eqref{NSE} leads to a coupled set of equations governing the radial structure of the disk. In the present context, these equations are solved only within the physically admissible radial domain identified from the geodesic analysis. The surface density is defined by the vertical integration of the rest mass density,
\begin{align}\label{sigma2}
\Sigma=\int_{-H}^{+H}\rho\,dz \simeq 2\rho H,
\end{align}
where $H(r)$ is the half thickness of the disk. For steady accretion, the mass accretion rate is
\begin{align}\label{mdot}
\dot{M}=-2\pi r\,\Sigma\,u^r=\mathrm{const.},
\qquad
u^r=-\frac{\dot{M}}{2\pi r\,\Sigma},
\end{align}
where $u^r<0$ is the radial inflow velocity.

In the thin disk approximation, the heat flux is assumed to be purely vertical. The radiative energy flux emitted from each face of the disk therefore satisfies \citep{1973blho.conf..343N,1974ApJ...191..499P}
\begin{align}
q^z(r,z)=F(r)\,\frac{z}{H(r)}.
\end{align}
Using Equations~\eqref{restmasscon}, \eqref{energycon}, and \eqref{NSE}, one then obtains the relativistic flux formula
\begin{align}\label{ene}
F(r)=-\frac{\dot{M}\,\Omega_{,r}}{4\pi\sqrt{-g}\,[E-\Omega L]^2}
\int_{r_0}^{r}(E-\Omega L)\,L_{,r}\,dr,
\end{align}
where $E$, $L$, and $\Omega$ are the specific energy, specific angular momentum, and angular velocity of equatorial circular motion, respectively, and $r_0$ denotes the inner edge of the radiating disk configuration.

The vertically integrated viscous stress is defined by
\begin{align}\label{w}
W=\int_{-H}^{+H}S^{\hat r\hat\phi}\,dz \simeq 2H\,S^{\hat r\hat\phi}
=2\alpha_{\rm vis}PH,
\end{align}
where $\alpha_{\rm vis}$ is the dimensionless viscosity parameter in this model, and $P$ is the total pressure at the disk midplane. The locally generated dissipative flux can then be written as
\begin{align}\label{navi}
F
\simeq-\sigma_{\hat r\hat\phi}\,W.
\end{align}
Assuming local blackbody emission and radiative diffusion in the vertical direction, the energy-transport equation takes the form
\begin{align}\label{OD}
F=\frac{8ac\,T^4}{3\kappa\Sigma},
\end{align}
or, equivalently $aT^4=(3\kappa\Sigma F)/8c\,$, where $\kappa$ is the Rosseland-mean opacity. We adopt
\begin{align}
\kappa=
0.40+0.64\times10^{23}
\left(\frac{\rho}{{\rm g\,cm^{-3}}}\right)
\left(\frac{T}{{\rm K}}\right)^{-7/2}
{\rm cm^2\,g^{-1}},
\end{align}
where the first term corresponds to electron-scattering opacity and the second to free-free absorption opacity. Here $a$ is the radiation density constant, related to the Stefan--Boltzmann constant $\sigma_{\rm SB}$ by
\begin{align}
a=\frac{4\sigma_{\rm SB}}{c}.
\end{align}
The total pressure is taken to be the sum of gas and radiation pressure,
\begin{align}\label{P}
P=\frac{\rho k_{\rm B}T}{m_p}+\frac{a}{3}T^4,
\end{align}
where $k_{\rm B}$ is Boltzmann's constant and $m_p$ is the proton mass. Finally, vertical hydrostatic balance in the comoving frame gives
\begin{align}\label{VP}
\frac{P}{\rho}=\frac{1}{2}\frac{(HL)^2}{r^4},
\end{align}
which follows from the relativistic Euler equation without further approximation \citep[equation 28]{1997ApJ...479..179A}.

\begin{figure*}
    \centering 
     \includegraphics[width=0.9\textwidth]{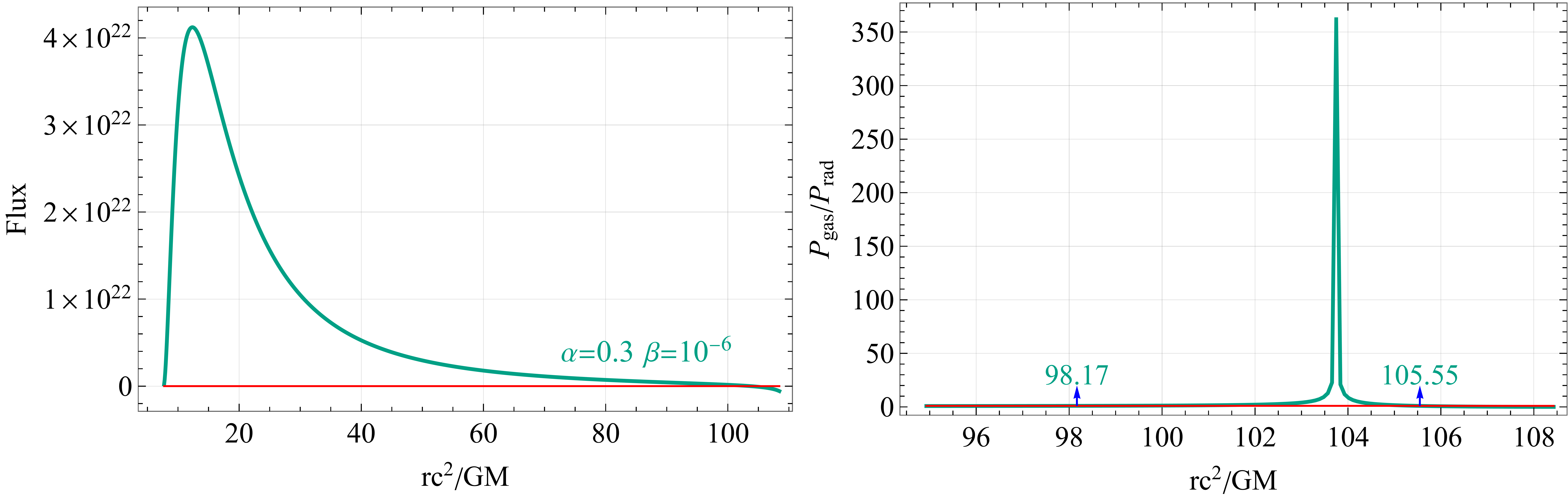}
      \includegraphics[width=0.9\textwidth]{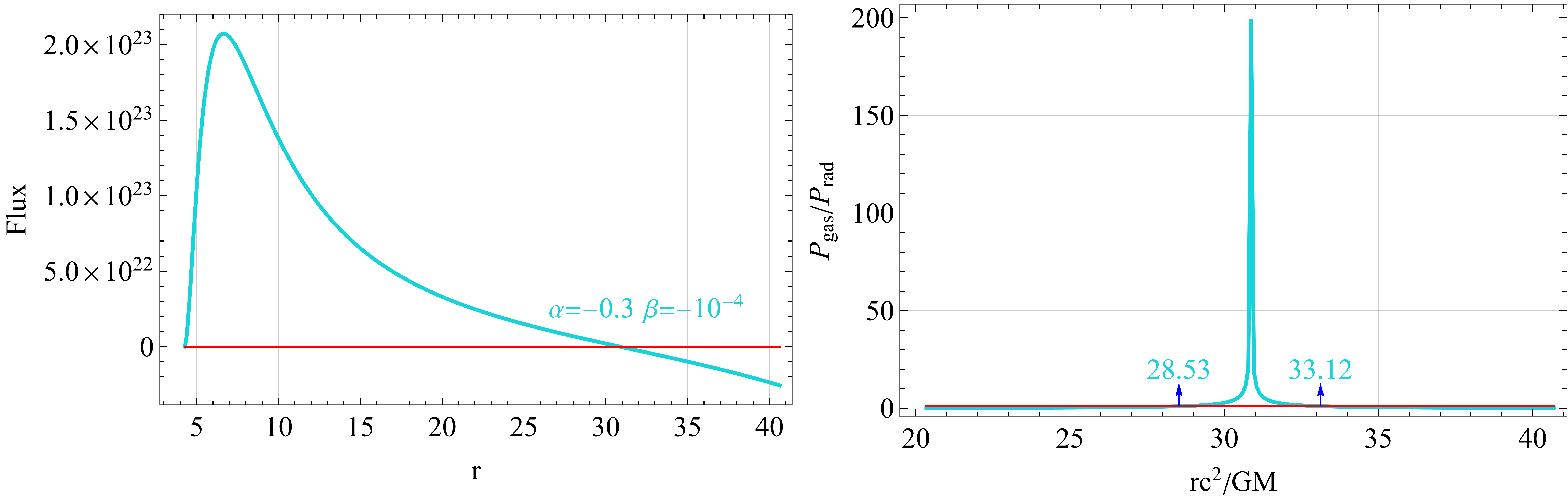}
      \includegraphics[width=0.9\textwidth]{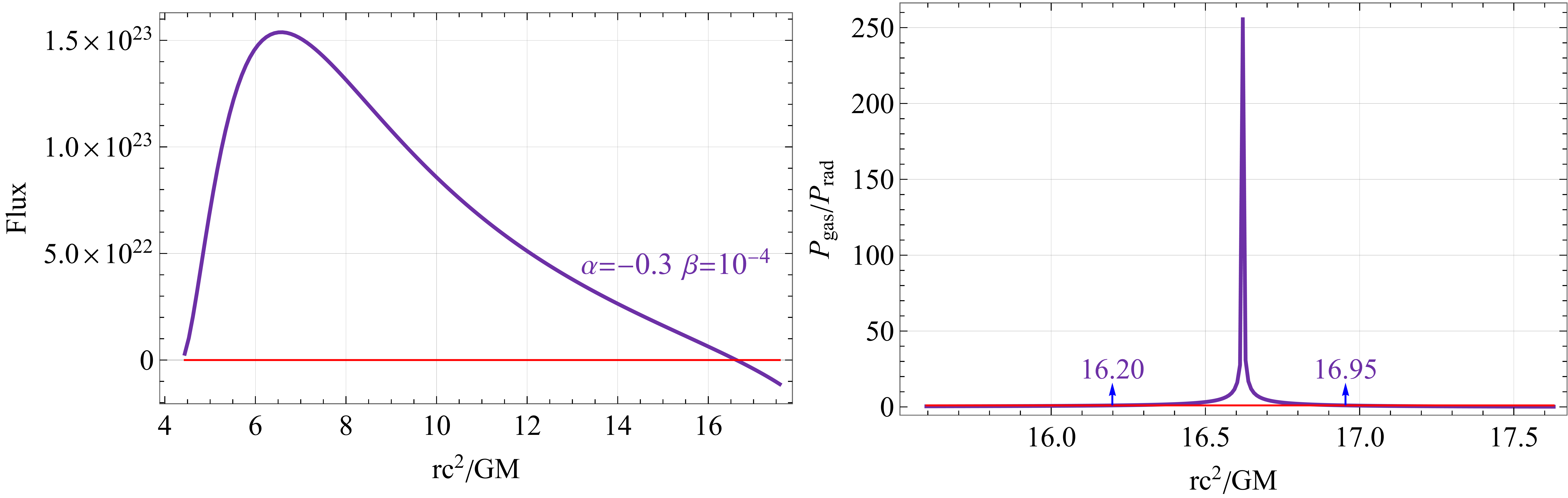}
    \caption{\label{fig:flux_ratio}Thin disk diagnostic panel. Left panels: radiative flux $F(r)$ for three representative distorted configurations. Right panels: the corresponding pressure ratio $P_{\rm gas}/P_{\rm rad}$, with the horizontal red line indicating $P_{\rm gas}=P_{\rm rad}$. In contrast to the asymptotically flat seed cases, for nonvanishing $\beta$ the flux vanishes at a finite radius $r_{\rm F}$. The pressure ratio exhibits a sharp excursion in the same outer region and crosses unity at two radii, $r_{\rm eq}^{\rm in}$ and $r_{\rm eq}^{\rm out}$. In all displayed cases, the outer equality radius $r_{\rm eq}^{\rm out}$ coincides numerically with the zero-flux radius $r_{\rm F}$, indicating that the radiating thin disk solution terminates at a finite outer boundary set by the local distorted geometry. }
\end{figure*}

\subsection{Flux termination and the gas-radiation pressure transition}\label{subsec:flux}

In the standard stationary, optically thick thin disk model, the vertically averaged structure equations admit different local branches according to which contribution dominates the pressure and which opacity source controls the radiative transport \citep{1973blho.conf..343N,1974ApJ...191..499P}. This classification is not merely formal: the transition radii between the branches determine how the local solutions are matched into a global accretion configuration. In the present spacetime, this point is particularly important, because the disk may be truncated by the finite physically meaningful domain of the local distorted geometry. As a result, not all of the standard radial regions need be realized. The same local classification also applies when the radiating structure is annular rather than a continuous disk.

Within this framework, three canonical regimes can be distinguished:

\begin{enumerate}[label=\Roman*.]

\item Inner region: radiation-pressure / electron-scattering dominated.
In the innermost part of the radiating flow,
\begin{align}
P \simeq P_{\rm rad}, \qquad \kappa \simeq \kappa_{\rm es}.
\end{align}
This regime is controlled by radiation pressure, while electron scattering provides the dominant opacity. Because it is closest to the inner edge of the radiating configuration, it is also the region most sensitive to the strong field geometry and to the boundary conditions.

\begin{figure*}
    \centering \includegraphics[width=0.95\textwidth]{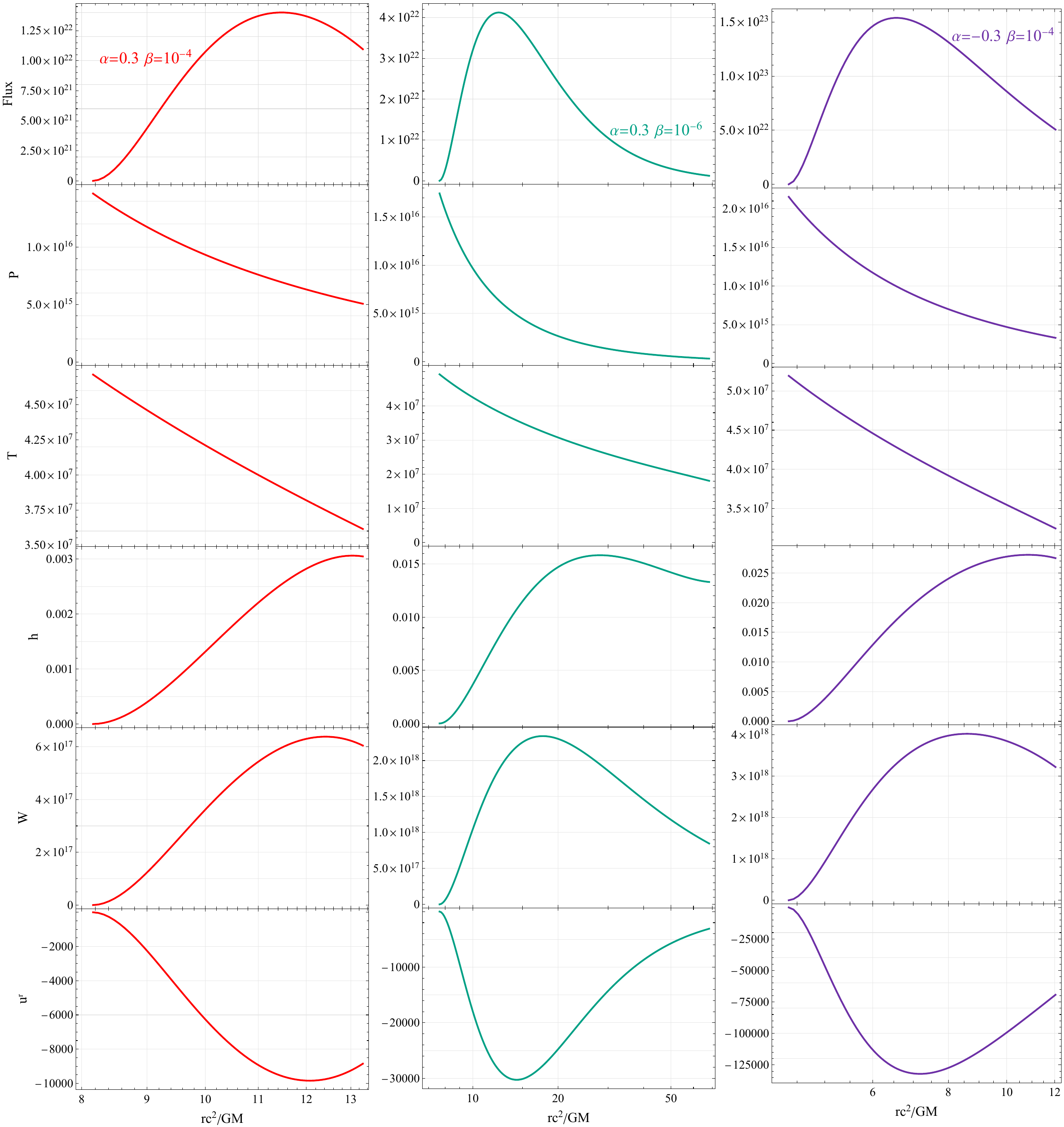}
    \caption{\label{fig:ringpropertiespositive} Representative radial profiles of ring-like thin accretion solutions for positive quadrupolar distortion $\beta$. From top to bottom, the panels show the emitted flux $F$, pressure $P$, temperature $T$, relative thickness $h$, vertically integrated viscous stress $W$, and radial velocity $u^r$ in the SI units. Each column corresponds to a different set of spacetime parameters, as indicated in the panels. The first two columns compare different values of the deformation parameter $\alpha$ at fixed positive $\beta$, while the last two compare different values of $\beta$ at fixed $\alpha$. In all cases, the radiating configuration is confined to a finite annular interval, consistent with the existence of a ring rather than a globally extended disk. The pressure and temperature decrease outward, the relative thickness remains small, the flux and stress peak within the annulus, and the radial velocity remains negative, as expected for inward accretion. Increasing positive $\beta$ makes the annulus more compact, in agreement with the reduction of the stable ring size inferred from the marginal stability analysis.}
\end{figure*}

\item Middle region: gas-pressure / electron-scattering dominated. 
At somewhat larger radii, the pressure becomes gas dominated while the opacity is still primarily due to electron scattering:
\begin{align}
P \simeq P_{\rm gas}, \qquad \kappa \simeq \kappa_{\rm es}.
\end{align}
This intermediate branch connects the inner radiation-supported solution to the outer, cooler part of the flow.

\item Outer region: gas-pressure / free-free-opacity dominated.  
Farther out, the disk remains gas-pressure dominated, but the opacity is now mainly governed by free-free absorption:
\begin{align}
P \simeq P_{\rm gas}, \qquad \kappa \simeq \kappa_{\rm ff}.
\end{align}
This is the standard outer branch of the thin disk solution.

\end{enumerate}
The boundaries between adjacent regions are determined by the conditions
\begin{align}
P_{\rm gas}=P_{\rm rad},
\qquad
\kappa_{\rm es}=\kappa_{\rm ff},
\end{align}
together with the structure equations of the previous subsection. In the asymptotically flat case, these local branches can be patched smoothly to construct a global disk solution. In the present distorted geometry, however, the admissible radiating region may terminate before one or more of these transition radii are reached. For this reason, the locations of the pressure- and opacity-transition radii provide an additional diagnostic of the physically meaningful extent of the disk, and later of the ring-like configurations that may arise in the distorted spacetime.

A particularly important consequence of the local character of the distorted spacetime appears in the radiative flux profile. In the asymptotically flat seed geometries, namely Schwarzschild and the undistorted q-metric with $\beta=0$, the thin disk flux remains positive throughout the admissible disk region and tends to zero only asymptotically as $r\to\infty$. Once the external distortion is switched on; however, this behavior changes qualitatively. Because the flux is proportional to $\Omega_{,r}$ through \autoref{ene}, a finite radius extremum of the angular velocity implies the existence of a finite radius $r_{\rm F}$ at which
\begin{equation}\label{frf=0}
F(r_{\rm F})=0.    
\end{equation}
If the disk solution be extended beyond this point, the formal flux would become negative, which is not physically acceptable for a radiating thin disk. We therefore interpret $r_{\rm F}$ as the outer boundary of the physically meaningful radiating region associated with the local distorted geometry.

This behavior is illustrated in \autoref{fig:flux_ratio}. The left panels show that, for nonvanishing $\beta$, the flux rises from the inner edge, reaches a maximum, and then decreases to zero at a finite radius. The right panels display the corresponding pressure ratio $P_{\rm gas}/P_{\rm rad}$. In each case, the ratio develops a sharp excursion near the same outer region and intersects the line $P_{\rm gas}/P_{\rm rad}=1$ at two radii,
\begin{equation}
r_{\rm eq}^{\rm in}<r_{\rm eq}^{\rm out}.
\end{equation}
The physically relevant point is that the outer equality radius $r_{\rm eq}^{\rm out}$ coincides numerically, within the resolution of the present calculation, with the zero-flux radius $r_{\rm F}$. In other words,
\begin{equation}
r_{\rm F}\simeq r_{\rm eq}^{\rm out}.
\end{equation}
This coincidence is significant. It shows that the finite-radius suppression of the radiative flux is accompanied by the loss of radiation-pressure dominance at the outer edge of the thin disk solution. Accordingly, the outer boundary of the physically meaningful disk can be identified simultaneously by two independent criteria: the vanishing of the flux and the outer solution of $P_{\rm gas}=P_{\rm rad}$. The narrow interval between $r_{\rm eq}^{\rm in}$ and $r_{\rm eq}^{\rm out}$ marks a transitional zone in which the standard hierarchy of the local disk branches is strongly distorted before the radiating solution terminates. This provides a thin disk manifestation of the finite radial domain of validity already inferred from the geodesic analysis.

In fact, the flux boundary is a thin disk diagnostic, i.e., is the correct bound for a steady, Keplerian, radiatively efficient, thin flow. However, the shear boundary that we discussed in \autoref{fig:shear} is closer to being a geometry based diagnostic of where the local solution starts to lose physical meaning.

To compare the outer radial bounds inferred from the shear  \autoref{shearsigmahat} and from the flux \autoref{ene}, we define
\begin{equation}
r_\sigma:\ \sigma_{\hat r\hat\phi}(r_\sigma)=0,
\end{equation}
also we already specify $r_F$ with $F(r_F)=0$  \autoref{frf=0}. In the thin, nearly circular regime one has

\begin{equation}
\sigma_{\hat r\hat\phi}(r)\ \propto\ \Omega_{,r}(r)  \,\Rightarrow \,\sigma_{\hat r\hat\phi}=0 \Longleftrightarrow \Omega_{,r}=0,
\end{equation}
whereas the relativistic thin disk flux has the schematic dependence (cf. \autoref{ene}) that we can write it as
\begin{equation}\label{Ffactor}
F(r)\ \propto\ -\,\Omega_{,r}(r)\,\frac{\mathcal I(r)}{(E-\Omega L)^2}, \quad \mathcal I(r)=\int_{r_0}^{r}(E-\Omega L)\,L_{,r}\,dr .    
\end{equation}
Therefore, for a fixed parameter pair $(\alpha,\beta)$, we have 
\begin{equation}\label{OmegaImpliesFlux}
\Omega_{,r}=0 \ \Longrightarrow\ F=0,
\end{equation}
whereas the converse holds only when the transport factor does not vanish,
\begin{equation}\label{FluxImpliesOmega}
F=0\ \Longrightarrow\ \Omega_{,r}=0
\qquad \text{provided}\qquad {\cal I}(r)\neq 0.
\end{equation}
Accordingly, defining $r_\sigma$ by $\sigma_{\hat r\hat\phi}(r_\sigma)=0$ and $r_F$ as the first outer root of $F(r)$, one expects $r_F=r_\sigma$ whenever ${\cal I}(r_F)\neq0$. A genuine separation $r_F<r_\sigma$ can occur only if ${\cal I}(r)$ vanishes at some radius while $\Omega_{,r}\neq0$, which in practice requires nonstandard transport behavior such as a sign change of $L_{,r}$. Hence, these diagnostics are meaningful only within the orbital-admissibility domain selected by $N_L>0$, $N_E>0$ and $D>0$ given in \autoref{eq:ND}; therefore the conservative outer edge of a radiatively efficient thin disk realization is taken as
\begin{equation}
r_{\rm out}=\min\{r_{\rm orb},\,r_F\},
\end{equation}
where $r_{\rm orb}$ denotes the outer boundary of admissible circular motion.



\subsection{Properties of thin rings/disk for quadrupolar distortion}
\begin{figure*}
    \centering 
     \includegraphics[width=0.95\textwidth]{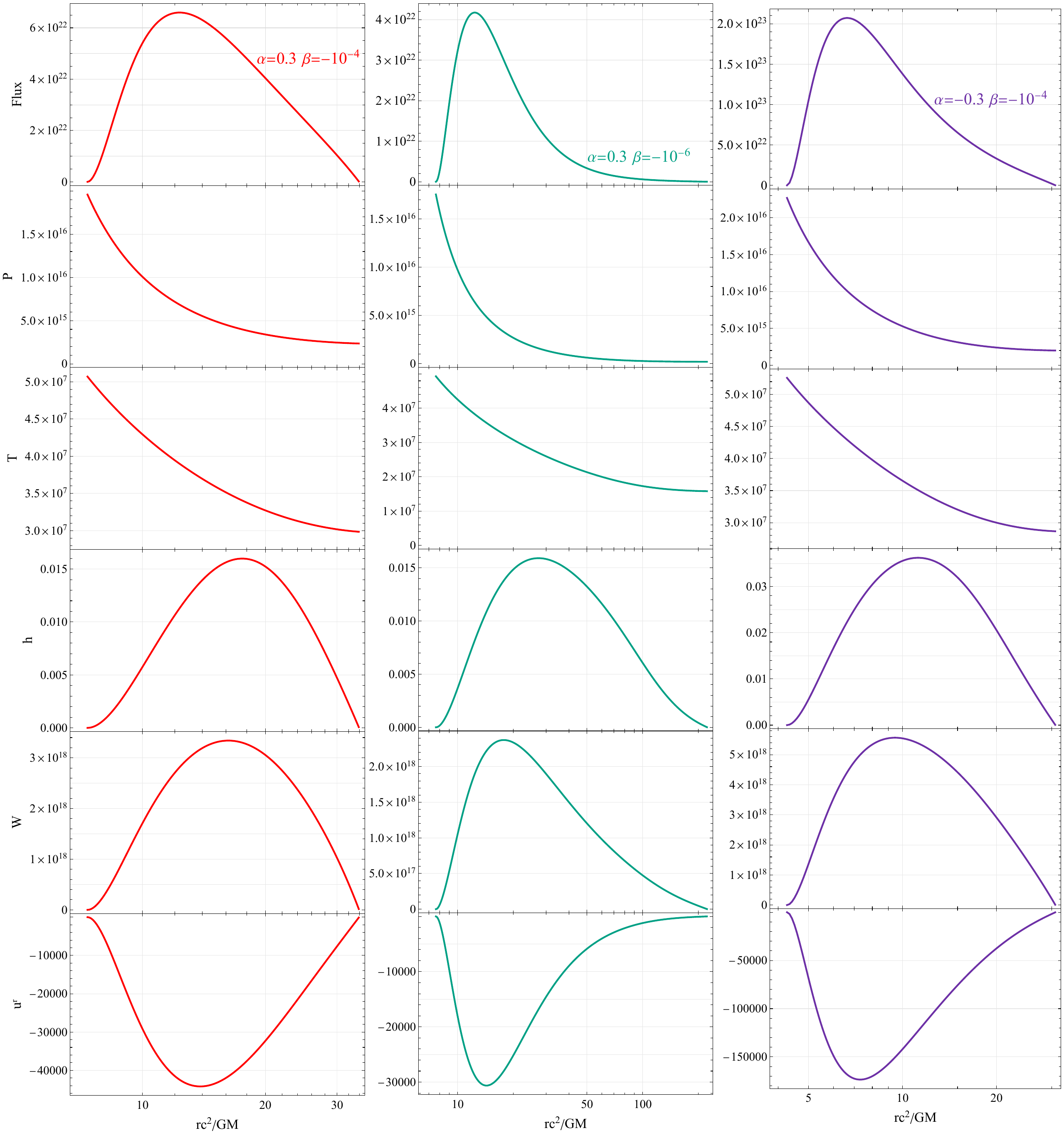}
    \caption{\label{fig:ringpropertiesnegative}Representative radial profiles of truncated thin disk solutions for negative quadrupolar distortion. From top to bottom, the panels show the emitted flux $F$, pressure $P$, temperature $T$, relative thickness $h$, vertically integrated viscous stress $W$, and radial velocity $u^r$ in the SI units. The columns correspond to $(\alpha,\beta)=(0.3,-10^{-4})$ (left), $(0.3,-10^{-6})$ (middle), and $(-0.5,-10^{-4})$ (right). In contrast to the positive $\beta$ case, where the radiating configuration becomes annular, negative $\beta$ yields a connected but finite thin disk. The outer cutoff radius increases strongly as $|\,\beta\,|$ decreases, while changes in $\alpha$ modify the location and extent of the disk without altering its overall  character of truncated disk.} 
\end{figure*}

In this subsection, we explore the radial profiles of the flux $F$, pressure $P$, temperature $T$, relative thickness $h$, vertically integrated stress $W$, and radial velocity $u^r$ for such ring-like/disk configurations, via solving the system of equations which presented in \autoref{eq:equations} semi-analytically. The results were presented in \autoref{fig:ringpropertiespositive} and \autoref{fig:ringpropertiesnegative}.
As we have already seen, for positive values of the distortion parameter $\beta$, the orbital analysis may yield a disconnected stable region, so that the radiating configuration is no longer a standard extended thin disk but a finite annulus. In that case, all hydrodynamical and radiative quantities are defined only on a bounded radial interval, and the resulting solution should be interpreted as a thin ring. 

In particular, \autoref{fig:ringpropertiespositive} shows several generic features. First, the pressure and temperature decrease outward across the annulus, while the relative thickness remains small throughout the entire domain. This confirms that the solutions remain geometrically thin even when the radiating structure is ring-like rather than globally disk-like. Second, both the emitted flux and the viscous stress develop a characteristic single-peaked profile inside the annulus, whereas the radial velocity remains negative, as expected for inward accretion, and reaches its largest magnitude away from the boundaries. These behaviors are consistent with a finite radiating flow confined between two radial stability boundaries.

The \autoref{fig:ringpropertiespositive} also shows how the ring properties respond to changes in the spacetime parameters. Comparing the first two columns, which correspond to the same positive $\beta$ but different values of $\alpha$, one sees that changing the intrinsic deformation primarily shifts the location and radial extent of the annulus, while preserving the overall form of the profiles. By contrast, comparing the last two columns, in which $\alpha$ is held fixed and $\beta$ is varied, shows that increasing the positive quadrupolar distortion compresses the annulus and moves it closer to the compact object. This is in full agreement with the earlier analysis of the marginally stable radii, which showed that positive $\beta$ reduces the size of the stable ring region.

It is important to note that, for ring solutions, the outer edge of the radiating domain is determined by the end of the admissible annular interval rather than necessarily by the zero-flux radius (\autoref{subsec:flux}). Consequently, although the flux always vanishes at the inner boundary imposed by the standard no-torque condition, it need not vanish at the outer edge unless the flux-termination radius lies inside the same admissible interval. This explains why the ring profiles remain perfectly consistent with the finite domain interpretation even when the flux is still nonzero at the outer boundary of the plotted region.

For negative values of the quadrupolar distortion parameter, the admissible accretion region remains connected. In this case, the local distorted geometry supports a finite truncated thin disk rather than an annular ring. \autoref{fig:ringpropertiesnegative} identifies different features, as follows. The flux vanishes at both ends of the plotted interval: the inner zero is associated with the usual no-torque condition at the inner edge of the radiating flow, whereas the outer zero reflects the finite-radius flux termination induced by the local character of the distorted spacetime. The pressure and temperature decrease monotonically outward, while the relative thickness remains small throughout the entire configuration, confirming that the solutions remain within the thin disk regime. The viscous stress develops a single broad maximum inside the disk and tends to zero at both boundaries, and the radial velocity remains negative everywhere, as expected for inward accretion.

Also, \autoref{fig:ringpropertiesnegative} also makes clear that the negative $\beta$ configurations are substantially more extended than the ring-like solutions found for positive $\beta$. In particular, for fixed $\alpha=0.3$, reducing the magnitude of the distortion from $\beta=-10^{-4}$ to $\beta=-10^{-6}$ pushes the outer cutoff to much larger radius and broadens the disk significantly, while preserving the overall shape of the profiles. By contrast, comparing the first and third columns, which are computed at the same $\beta=-10^{-4}$ but for different values of $\alpha$, shows that the intrinsic deformation mainly shifts the location, width, and amplitude of the profiles without changing their qualitative behavior. Thus, for $\beta<0$, the external distortion produces a connected but finite thin disk configuration, whose radial extent is controlled primarily by $|\,\beta\,|$ and further modulated by $\alpha$.

Taken together, \autoref{fig:ringpropertiespositive} and \autoref{fig:ringpropertiesnegative} show that the sign of the quadrupolar distortion controls the global morphology of the accreting configuration: positive $\beta$ favors finite annular rings, whereas negative $\beta$ leads to connected but truncated thin disks.


\section{Summary and conclusions}\label{sec:summary}

In this work, we investigated the local distorted extension of the q-metric in order to determine the radial domain over which it can be regarded as a physically meaningful description of a deformed compact object embedded in an external quadrupolar environment. By construction, the external distortion destroys global asymptotic flatness, so our main question is where its local character remains self-consistent and astrophysically relevant. Our analysis shows that this question can be answered by combining orbital dynamics with the structure of thin accretion flows.

We first studied the behavior of the fundamental orbital quantities associated with equatorial circular timelike motion, namely the specific energy, the specific angular momentum, and the angular velocity. Their behavior already indicates that the distorted geometry admits a finite physically admissible region. In particular, while the seed asymptotically flat configurations display the standard monotonic behavior expected at large radii, the inclusion of the external quadrupolar field may induce an extremum in the angular velocity at a finite distance from the compact object. This indicates that the influence of the external matter becomes dynamically relevant and that the local spacetime description should not be extrapolated arbitrarily far from the compact object.

By using effective potential and of the marginally stable circular orbits we found that this distorted geometry develops more complicated stability structure compare to q-metric or Schwarzschild solution and, depending on the values of $\alpha$ and $\beta$, may admit an additional outer marginally stable branch. This is especially important for positive values of the external quadrupolar parameter, for which the stable circular region can become annular. In that case, the geometry naturally supports ring-like accretion configurations bounded by two marginally stable radii. For negative $\beta$, by contrast, the stable region remains connected but terminates at a finite outer radius, leading instead to truncated thin disk solutions. The sign of the external quadrupolar distortion therefore controls the global morphology of the accreting configuration.

We then showed that thin disk observables provide independent and physically more stringent diagnostics of the same finite validity domain. The local orthonormal shear component remains negative throughout the admissible region, as expected for inward accretion and outward angular momentum transport, but for nonvanishing distortion it may approach zero at a finite radius rather than only asymptotically. This identifies a second, fully local marker of the outer boundary of the physically meaningful flow. An equally important result follows from the radiative flux. In the distorted spacetime, the flux can vanish at a finite radius, beyond which the formal thin disk solution would cease to be physically acceptable. Within the explored parameter range and numerical resolution, this zero-flux radius coincides with the outer location at which gas pressure and radiation pressure become equal. This coincidence is highly significant, because it shows that the termination of the radiating solution is simultaneously encoded in the orbital structure, the shear profile, and the thermodynamic state of the disk.

The resulting accretion solutions remain internally consistent within the allowed radial interval. In both the truncated disk and ring-like cases, the pressure and temperature decrease outward, the relative thickness remains small, the viscous stress develops the expected peaked profile, and the radial velocity retains the sign appropriate for inward flow. Thus, the standard thin disk assumptions remain applicable inside the domain selected by the distorted geometry, even though the background spacetime itself is only local. This makes the analysis physically meaningful: the finite extent of the admissible accretion region is not an artifact of a single diagnostic, but the outcome of several mutually consistent and independently derived criteria.

This gives a physically motivated and observationally relevant meaning to the notion of a local exact solution. At the same time, the analysis demonstrates that an external quadrupolar environment can alter the structure of accretion in a qualitatively nontrivial way, producing either finite rings or truncated disks and shifting the relevant boundaries of the flow far from those of the seed spacetime. The next generalization of this work will be considering rotation into the system and explore the effect of magnetic fields on the shape and stability of the valid domain.

\begin{acknowledgments}
Authors gratefully acknowledge Ramesh Narayan for insightful discussions that contributed to the initiation of this work. This research is supported by the University of Waterloo. Also, Natural Sciences and Engineering Research Council of Canada, by the Government of Canada through the Department of Innovation, Science and Economic Development and by the Province of Ontario through the Ministry of Colleges and Universities at Perimeter Institute. In addition, thanks to the research training group GRK 1620, Models of Gravity, funded by the German Research Foundation (DFG).   
\end{acknowledgments}

\bibliographystyle{plainnat}
\bibliography{pq-diskbib}
\end{document}